\newif\ifShowKeys
\ShowKeystrue
\newif\ifshowtikz
\showtikztrue
\showtikzfalse   % <---- comment/uncomment that line
\documentclass[11pt]{article}
	\pdfoutput=1
\usepackage[T1]{fontenc}

\usepackage{listings}
\lstnewenvironment{arkady}  % usage \begin{arkady} ... \end{arkady}
{\lstset{language=C,frame=trbl,basicstyle = \footnotesize \ttfamily , breaklines = true,showstringspaces=false}}{}

\usepackage{datetime}
\usepackage{comment}					% to comment large parts of text

\usepackage[usenames,dvipsnames]{xcolor}
\usepackage[setpagesize=false,pagebackref=false, 
linktocpage, bookmarksopen=true, colorlinks=true, 
linkcolor=Maroon,citecolor=Maroon,urlcolor=Maroon]{hyperref}

\usepackage[parsep]{collref}				% collect references in groups

\usepackage{amsmath, amssymb,amsthm}
\usepackage{stackrel}
\numberwithin{equation}{section}
\usepackage{bm,environ,mathrsfs,array,arydshln}
\usepackage{booktabs,float,slashed}
\usepackage{appendix}
\usepackage[mathcal]{euscript}
\usepackage{tensor} 						% Ratcliffe package to write tensors
\usepackage{mathabx}
\usepackage[vcentermath]{youngtab}
\usepackage{simpler-wick}

\usepackage{graphicx,epsfig,epic}
\usepackage{tikz}
\usepackage{tikz-feynman} 
\allowdisplaybreaks

\usepackage{framed}						% for shaded equations \begin{shaded} + \end{shaded} or \bs + \es
\definecolor{shadecolor}{rgb}{0.9996078, 0.984314, 0.960784}
\definecolor{framecolor}{rgb}{0,0,0}
\definecolor{TFTitleColor}{RGB}{1,1,1}
\newenvironment{frshaded}{%
    \MakeFramed {\FrameRestore}}%
    {\endMakeFramed}

\definecolor{myred}{RGB}{233, 33, 45}

\newcommand{\bs}{\begin{frshaded}}			% framed with background in shadecolor 
\newcommand{\es}{\end{frshaded}\noindent}

\def\ba#1\ea{\begin{align}#1\end{align}}		        %  clever way to bypass the known problem...
\newcommand{\be}{\begin{equation}}
\newcommand{\ee}{\end{equation}}
\newcommand{\bea}{\begin{equation} \begin{aligned}} 
\newcommand{\eea}{\end{aligned} \end{equation}}
\newcommand{\mc}{\mathcal }
\newcommand{\wh}{\widehat}
\newcommand{\wt}{\widetilde}
\newcommand{\mb}{\mathbb }
\newcommand{\mk}{\mathfrak}
\newcommand{\la}{\label}
\newcommand{\eps}{\varepsilon}

\newcommand{\lp}{\notag \\ & }

\DeclareMathOperator{\Tr}{\text{Tr}}
\DeclareMathOperator{\tr}{\text{tr}}
\DeclareMathOperator{\vol}{vol}

\newcommand{\N}{\mathcal N}

\def \AdS  {{\rm AdS}}
\def \adst {AdS$_3$}  \def \ZZ  {{\mathbb Z}}

\renewcommand{\l}{\lambda}

\newcommand{\adstwo}{AdS$_{2}$ }

\newcommand{\LDR}{\ell}
\newcommand{\sss}{{{\rm G}_\theta}}
\newcommand{\gs}{{\rm G}}
\newcommand{\G}{\Gamma}
\newcommand{\w}{{\rm w}}

\def \np {\newpage}
\def \ed {\np \small
\baselineskip 11pt
\bibliography{BT-Biblio}
\small
\bibliographystyle{JHEP-v2.9}
\end{document}
}
\def \iffa  {\iffalse}
\def \te {\textstyle}
\newcommand{\rf}[1]{(\ref{#1})}
\def\ov{\over}
\def \ci {\cite}
\def \foot {\footnote}

\def\la{\label}\def \a {\alpha}
\def\foot{\footnote}
\def \adss {AdS$_5 \times S^5$\ }

\def \s {\sigma} \def \del {\partial} 
\def \ha {\tfrac{1}{2}}
\def \DD {{\rm D}}
\def \bb {{\rm b}}
\def \OO {{\cal O}}
\def \cc {{\rm c}}
\def \no {\nonumber}
  
 \def \g {\gamma}
 
\newcommand{\ad}{\text{ad}}

\newcommand{\Q}{\mathrm{Q}}

\newcommand{\kil}{\mathrm{kil}}

\newcommand{\mx}{|\vec{x}|}
\newcommand{\my}{|\vec{y}|}

\newcommand{\Pc}{\mathrm{P}}
\newcommand{\id}{\mathbb{I}}
\newcommand{\hyp}{\mathrm{H}}

\newcommand{\Df}{\mathfrak{D}}

\newcommand{\Gst}{\Gamma_{*}}

\begin{document}
\begin{titlepage}
\begin{tabbing}
\hspace*{10.5cm} \=  \kill % set the tabbings
\>  Imperial--TP--2025--SK--03 \\
\> %none
\end{tabbing}

%\centerline{\small\today\ -- \currenttime}

\vspace*{05 mm}
\begin{center}
{\Large\sc  \bf     % Subleading strong-coupling term in
%Subleading  term in 2-defect anomaly in (2,0) theory from  
%2-loop   %AdS$_3$  
2-loop   free energy of  M2 brane   in  AdS$_7 \times S^4$  %:  free energy on AdS$_3$  at 2 loops  
 \\ \vskip 6pt
and  surface defect  anomaly in (2,0) theory }%\vskip 9pt
%{\Large\sc      Notes}
\vspace*{07mm}

M. Beccaria$^{a}$, \ \ S.A. Kurlyand$^{b}$,\ \ \ A.A. Tseytlin$^{b,}$\footnote{Also at  ITMP of MSU  and Lebedev Inst.}

\vspace*{2mm}
{\small
	
${}^a$ Universit\`a del Salento, Dipartimento di Matematica e Fisica \textit{Ennio De Giorgi},\\ 
		and INFN - sezione di Lecce, Via Arnesano, I-73100 Lecce, Italy
			\vskip 0.1cm
${}^b$ Abdus Salam Centre for Theoretical Physics,\\ Imperial College London,  SW7 2AZ, U.K.
			\vskip 0.01cm
\vskip 0.2cm {\small E-mail: \texttt{matteo.beccaria@le.infn.it,\ s.kurlyand23@ic.ac.uk, \ tseytlin@ic.ac.uk}}
}
\vspace*{1.2cm}
\end{center}
\begin{abstract}  
	{A $\frac{1}{2}$-BPS surface operator  viewed as a conformal defect  in  6d (2,0) theory theory associated with $N$
	 coincident M5 branes is expected  to have 
	 a holographic description in terms of a probe M2 brane  wrapped on AdS$_3$ in  the AdS$_7\times S^4$  M-theory background. 
	 The M2 brane 
	  has the  effective tension ${\rm T_2}= {2\ov \pi} N$  so that  the large tension expansion corresponds to the $1/N$ expansion. 
	The   value  of the  defect conformal anomaly coefficient in $SU(N)$  (2,0)  theory 
	 was previously argued 
	  to be  b$=12N- 9 - 3N^{-1}$. \
	  %v4
	  At the same time, one may expect that  the probe M2  brane  ending on a stack of $N$  M5 branes 
	   should  represent  a Wilson  surface operator in the $U(N)$   rather than $SU(N)$   boundary 6d CFT, and in this case 
	    one should   get   b$=12N- 9 $, i.e.  the $N^{-1}$ term  (that in the $SU(N)$ expression  ensures that b  vanishes for $N=1$)  should be absent. 
	%v4
	By semiclassically quantizing M2 brane,  it  was found in arXiv:2004.04562 that the  first two terms  in b  are indeed reproduced  by the classical  and 1-loop 
	corrections  to  the M2  free energy. 
	Here we address  the question  of  the    value of the  next    2-loop  term
	 in the M2 brane free energy,  i.e.  the coefficient of the  $N^{-1}$  term in b. 
	 %%%%%%%%%%%%%%%%%%%%%%%%%%%%
	 Remarkably, despite  the  general  non-renormalizability of the  standard  BST M2 brane  action  we find that
	 the 2-loop correction  to the  free energy  of the AdS$_3$  M2 brane in AdS$_7\times S^4$ 
	 is UV finite (modulo  power divergences that  can  be removed by 
	  an analytic regularization).  Moreover,   the 2-loop correction     vanishes  in  both 
	   dimensional   and  $\zeta$-function regularizations. 
	   This supports the interpretation that the M2-brane probe computation
	    captures the surface-defect anomaly of the $U(N)$, rather than the $SU(N)$, boundary 6d  theory.	  } \end{abstract}
\vskip 3.5cm
		%Keywords: {\sc insert here keywords}	%}
	%	\centerline{\small\today\ -- \currenttime}
\end{titlepage}
%%%%%%%%%%%%%%%%%%%%%%%%%%%%%%%%
%%%%%%%%%%%%%%%%%%%
%\end{comment}
%%%%%%%%%%%%%%%%%%%%%%%%%%%%%%%%%%%%%%%%%%%%%%%%%%%%%%%%%
{\small
\makeatletter
\newcommand*{\toccontents}{\@starttoc{toc}}
\makeatother
\toccontents
}
%\tableofcontents
%\vspace{1cm}
\def \adst {AdS$_2$ }
\def \ept {\eps'}
\def \adstwo {\adst}
  \def \rmb {{_{\rm b}}}   \def \rmf {{_{\rm f}}}  \def \rmbf {{_{\rm bf}}}
 \def \DD {{\mk D}}  \def \b {\beta } 
\def \m {\mu} \def \n {\nu} \def \L {\mc L}
 \def \ep {\eps}
  \def \ads {{\rm AdS}} \def \four {\tfrac{1}{ 4}}
\def \fo {\tfrac{1}{4}} \def \beg {\be}
\def \F  {F} \def \RR {{\Lambda_{\rm IR}}} \def \TT {{\rm T}}
\def \DD {{\rm D}}
\def \TT  {{\rm T} } %{_{\rm M2}}   }
\def \ep {\epsilon}
\def \vd {\dot v}   \def \ad {\dot a} \def \zb {\bar z} 
\def \RR {{\bar \Lambda }} 

\def \iffa {\iffalse} \def \aa {{\rm a }}  
\def \Tr {{\rm Tr}}
\def \tr {{\rm tr}}
\def \vol   {{\rm vol}}
\def \rr {{\rm r}}
\def \OO {{\cal O}}
\def \cc {{\rm c}}
\def \kk {\kappa}\def \ka {\kappa} \def \L  {\Lambda}
\def \rT {{\rm T}}
\def \tt {{\rm t}}
\def \half {\tfrac{1}{2}}
\def  \adsss  {AdS$_7\times S^4\ $}
\def  \adssp  {AdS$_5\times S^5\ $}
\def \adss {AdS$_7\ $}
\def \adst   {AdS$_3\ $}
\def \N   {{\cal N}}
\def \half {\tfrac{1}{2}}
\def \xx {{ x}}
\def \vx {\vec{x}}
\def \fo {{1\ov 4}}\def \n {\nu}
\def \la {\label} \def \fo {{1\ov 4}}
\def \ed   {\end{document}}
\def \beg  {\begin{equation}}
\def \eeg {\end{equation}}
\def \tet {\textstyle}
\def \lag {\langle} \def \rag {\rangle}
\def \xxx {z}
\def \xxxx {{\rm z}}
\def \ze {\chi}
\def \zz {{v}}
\def \adsp {AdS$_5 \times S^5$}
\def \rh   {\hat {\rm G}}
\def \rG {{\rm G}}
\def \rT {{\rm T}} 
\def \ssss   { {\hat {\rm G}}_\theta}
\def \z  {\zeta} 
\def \rhg {\tilde {\rm G}}
\def \w  {\s}
\def \halpha {{\hat \alpha}}  \def \hbeta {{\hat \beta}} \def \hgamma {{\hat \gamma}}
\def \K {{\rm K}}
\def \gs  {\rG}   \def \rh {\hat \rG}
\def \tgt {\tilde {\rm G}_\theta}
\def \bgt {\bar {\rm G}_\theta} 
\def \La  {\Lambda}  \def \tG {G_\theta}  
\def \hyp {{\rm H}} \def \ad {{\rm ad}\, }
\def \Xi {\varkappa}
\def \onevec  {{\bf 1}}

\setcounter{footnote}{0}

\section{Introduction \la{sec1}}

%
%A different  case is M2  brane wrapped on AdS$_3$ in AdS$_{7}\times S^4$  with 2-loop 
%should reproduce  $1/N$ term  in the 2-surface  boundary conformal anomaly. 
%\ci{Beccaria:2026ffm}  \ci{Beccaria:2025yyy}

It was suggested in   \ci{Drukker:2020swu}  that   considering 
 a  quantum  M2 brane  wrapped on $\ads_3\subset \ads_7$ in  $\ads_7 \times S^4$  
background   one may   capture  not only the leading in $N$  but also subleading coefficients in the 
defect b-anomaly   coefficient of the $S^2$   conformal  defect in the  boundary theory
 \be \la{01}
\bb  = {12 }  N -  9 + \OO(N^{-1}) \,. \ee
  % (see \rf{27})
  Here $N$ is the number of M5 branes forming the $\ads_7 \times S^4$  
  background. % (rank of the  (2,0)   boundary  CFT). 
  It is   related to the effective  dimensionless  M2  brane  tension  as 
  ${\rT}_2={2\ov \pi} N$.
  The  first term in \rf{01}   corresponds to  the  classical M2  action contribution 
   while  the second one is the 1-loop  contribution   to the $\ads_3$ 
  M2  free energy $F$   \ci{Drukker:2020swu}.\foot{For a general   discussion  
  of the leading   brane-probe  action contribution to a boundary  defect anomaly see   \ci{Graham:1999pm}. 
    For an arbitrary  2-surface  a defect operator has three anomaly coefficients,%
each multiplying a particular conformally invariant integral on the surface related to its topology, 
extrinsic curvature and background Weyl tensor \cite{Schwimmer:2008yh}. }

% Eq.\rf{01}  is   consistent   with the expression for the b-anomaly coefficient 
%    found  \cite{Estes:2018tnu,Jensen:2018rxu}
 %   from the entanglement entropy  computation for the 
%``bubbling''   \ci{DHoker:2008lup} M5-M2 geometry

%\foot{
The general expression for  the b-anomaly corresponding to 
   a   $1\ov  2$-BPS  surface defect 
operator in (2,0)  theory   which  corresponds to a
%v4
%  $SU(N)$ 
  representation of group $G$  with the  Young tableau   having  a large number  of boxes is  given by 
  \cite{Estes:2018tnu,Jensen:2018rxu,Chalabi:2020iie}
\be \la{001}
\bb= 24(\rho, \lambda) +3 (\lambda, \lambda)  \ , \ee
 where 
 $\rho$ is the Weyl vector  %of  $SU(N)$ 
   and $\lambda$ is the  highest weight of the  % $SU(N)$ 
representation. If one  formally   assumes  that this  relation  is  valid for  a  finite number of boxes  then  
%for a single M2 brane  corresponding to the
 for a  surface operator in the fundamental representation of $G=SU(N)$  (with $(\rho, \lambda)= \ha({N-1)} , \  (\lambda, \lambda) = 1- N^{-1}$)
 %which should   be described   by a single M2 brane probe 
  one  
  %v4
   finds\foot{Assuming $G=SU(N)$   the 
   expression  in  \rf{001}  and thus \rf{03}
 as an exact result in $N$  could  
   still  be   viewed as conjecture.  However, in \ci{Chalabi:2020iie} 
  a  similar expression for the  d$_2$   anomaly  coefficient  % (see \rf{1.8},\rf{H1}) 
 was derived as an exact result from a  superconformal index  computation. It also follows from the 5d Wilson-loop 
 localization computation as in  \ci{Mori:2014tca}. 
 Given that $ \bb$  and d$_2$   appear on an equal footing in the spherical entanglement
  entropy \ci{Jensen:2018rxu,Rodgers:2018mvq}
one may expect  %is inclined to believe
 that the expression for $\bb$  should   also  be exact. 
 Indeed,   the same  expression  for $\bb$   was found  on the dual CFT side 
  in \ci{Wang:2020xkc}  using  't Hooft anomaly  considerations.} 
  \be \la{03}
\bb\big|_{SU(N),\Box}  = 12N  ( 1 +\four {N^{-1}} )  (1-N^{-1})  =   12 N - 9  - {3N^{-1}}    \,. \ee
% }
The  first two terms  here  match the ones   in 
\rf{01}  while the $N^{-1}$  term should then   correspond to the 2-loop M2 brane contribution.

%v4
At the same time, one may argue  that the  M2  brane   probe ending on a stack of $N$  M5 branes   should 
instead  represent a  ``Wilson  surface''   operator in the  $U(N)$ (2,0) theory  ( understood as $A_{N-1}$ +  c.m. (2,0) tensor  multiplet) 
 rather than in the $SU(N)$ theory  \ci{vanMuiden:2026ffm}.
 This is  analogous to  the case of  the Wilson loop  in $\N=4$ SYM theory described  by a  fundamental open string
 ending on a stack of $N$ D3 branes  or  at the boundary of AdS$_5 \times S^5$  \ci{Maldacena:1998im}  (various discussions of  the $SU(N)$ \ci{Aharony:1998qu,Witten:1998wy,Aharony:2013hda,Kapustin:2014gua} versus  $U(N)$ \ci{Maldacena:2001ss,Belov:2004ht}  choice  have  appeared in \ci{Aharony:2016kai,Barbar:2025krh}).\foot{While   bulk AdS  observables  dual to closed string state  correlation functions 
 are not sensitive to the c.m. (singleton) $\N=4$   multiplet and thus  are dual to $SU(N)$   gauge theory correlators 
 the boundary ones  should  correspond to  the $U(N)$ gauge theory.  Then Chan--Paton index runs over all $N$ values,   
 i.e.  the boundary operator  should be dual to  the  trace in the fundamental representation of the full $U(N)$ group. 
The diagonal $U(1)$ may be free with respect to adjoint dynamics, but the external fundamental ``quark''  is charged under it.  Hence decoupling of the trace multiplet from interacting local correlators does not mean that its factor should be omitted  from the fundamental WL expectation value.
Explicitly,  for the $\ha$ BPS circular WL  one   finds \ci{Drukker:2000rr} that 
  $ \langle W\rangle_{U(N)}=
  \exp (\frac{\lambda}{8N^2})
  \langle W\rangle_{SU(N)} $.  The difference  is first visible at  $1/N^2$ or 1-loop (disk  with one  handle)     string  order
  (and is subleading   compared to   the leading strong-coupling contribution to  $ \langle W\rangle$  
   which is $\sim {\lambda^{3/2}\ov N^2}$).
  Checking directly that the  string   calculation in AdS$_5 \times S^5$  indeed  reproduces  $\langle W\rangle_{U(N)}$ rather than 
  $\langle W\rangle_{SU(N)} $ remains an outstanding challenge  at the moment.}

%   One may conjecture that the quantum  M2  brane free energy   should correspond to a surface defect in the $U(N)$ rather than $SU(N)$ 
  % boundary 6d  theory. 
%   While the  standard argument  \ci{Aharony:1999ti}  about decoupling of the $U(1)$ subgroup 
    %v4
   
   While  the  standard argument  \ci{Aharony:1999ti}  about decoupling of the $U(1)$ subgroup  may apply to   bulk   observables (captured by quantum M2 branes in the bulk of \adsss and, in particular, by 11d supergravity modes),   the boundary observables described   by  an M2 brane  ending on the AdS$_7$  boundary   should 
   correspond to defect  operators in the $U(N)$   (2,0)  theory that are not neutral with respect to $U(1)$. The general relation for $\bb$ in \rf{001} may be formally extended to the
$U(N)$ case.\foot{Note that in \ci{Estes:2018tnu} \rf{001} was abstracted
from their supergravity result {\it assuming} that it corresponds to the
$SU(N)$ case. For an $n$-symmetric $SU(N)$ representation
(in the context of \ci{Estes:2018tnu}, $N$ is the number of M5 branes and
$n$ is the number of M2 branes), one has from \rf{001}
$\bb=12nN-3n(4-n)-3n^2/N$. The supergravity discussion in
\ci{Estes:2018tnu} is applicable for $n,N\gg1$, with $n$ and $N$ being
of the same order. In this limit, the last $-3n^2/N$ term in $\bb$ is
subleading, and thus one cannot actually distinguish between the $SU(N)$
and $U(N)$ cases.} % (see also footnote \).}

Introducing an orthonormal basis $e_i$ of the dual of the maximal toral
subalgebra of $\mathfrak{u}(N)$, with the inner product induced by the trace
in the defining representation, the weight corresponding to the fundamental
representation may be written as
$\lambda_{U(N)}=e_1\equiv\lambda_{SU(N)}+N^{-1}\onevec$, where
$\onevec=\sum_{i=1}^N e_i$ and
$(\lambda_{SU(N)},\lambda_{SU(N)})=1-N^{-1}$.
Since the Weyl vector is traceless,
$(\rho,\lambda_{U(N)})=(\rho,\lambda_{SU(N)})=\frac{N-1}{2}$, while
$(\lambda_{U(N)},\lambda_{U(N)})=1$. One thus finds, instead of
\rf{03}, 
\be \la{002}
\bb\big|_{U(N),\Box}=12N-9 \, .
\ee
%\foot{\la{f8}
The absence of the $N^{-1}$ correction to the defect
anomaly coefficients $\bb$ and $d_2$, which have  similar structure, for a
general $(n,m)$ representation of $U(N)$ is implied also by the matrix-model
computation of the 5d Wilson-loop expectation value related to the $d_2$
coefficient \ci{Mori:2014tca}  (see appendix H, and in particular eq. (H.22),
in \ci{Beccaria:2024gkq} for details).

The  aim of the present   work is to  compute   the 2-loop   term   in the AdS$_3$  M2  brane  free energy. 
We will find that it vanishes,    in agreement with \rf{002}. 
%check if it  reproduces  the $ - {3N^{-1}} $   term in \rf{03}. 
Let us first review the general setup  in  \ci{Drukker:2020swu}. 
We  are going to  consider an      M2 brane  probe in the  AdS$_7 \times S^4$ background 
%which is the near horizon geometry of $N$ M5-branes
\begin{align}
 \la{452}
& ds^2 = L^2\big( ds^2_{_{{\rm AdS}_7}}+\rr^2 ds^2_{_{ S^4}}\big)
\ ,\qquad L^3 = 8 \pi N \ell_p^3 \ , \qquad  \rr=\tfrac12 \ , \\
& {\cal F}_4= dC_3=\tfrac{3}{8}  %  \pi^2
 L^3 \vol_{S^4}  \  , \qquad 
   \int_{S^4}   \vol_{S^4}= \vol(S^4) = \tfrac{8 \pi^2 }{3} \ . 
  \la{3101}
\end{align}  
% will use Euclidean signature   and 
The  BST    M2 brane    action \cite{Bergshoeff:1987cm,Bergshoeff:1987qx}
contains  a  bosonic part   (the standard Dirac-Nambu  term $S_1(X)$ and   a WZ-type term $S_2(X)$ 
describing the  coupling to the 3-form $C_3$)
and also  a fermionic  part  $S_f(X,\theta)$
\begin{align}
\la{31}
S=  &  S_b + S_f, \ \ \   S_b=S^{(1)}  + S^{(2)}\ , \  \ \ 
S^{(1)}  = T_2 \int d^3 \s \, \sqrt {h } \ , 
\qquad  h_{\m\n} = \del_\m X^M \del_\n X^N G_{MN} (X) \ , 
\\
S^{(2)}   &=  i T_2 \int d^3 \s \, \tfrac{1}{3!} \ep^{\m\n\l} C_{MNK}(X)\,  \del_\m X^M \del_\n X^N \del_\l X^K  \ ,  \la{3200} \qquad 
\qquad \te { T}_2=  {1\ov (2\pi)^2 \ell_p^3}\ . 
\end{align}
The  explicit  form of the M2  brane  action  in AdS$_{7}\times S^4$    can be  obtained  from   \ci{deWit:1998yu,Claus:1998fh}.
%(see also \ci{Sakaguchi:2003hk,Sakaguchi:2004bu}). 

The world-volume  geometry  of  an M2  brane
ending on  a  2-sphere  at the  boundary of AdS$_7$  is  described   by  the    (Euclidean) AdS$_3$  metric.
 The tree-level contribution  to the free   energy  $F=-\log Z$ 
  is  given by the  classical value of the  M2 brane action  which is   proportional 
to  the regularized  volume of the induced \adst metric 
 (cf. \cite{Berenstein:1998ij})\foot{Here $\RR= \Lambda_{_{\rm IR}} a$  where $\Lambda_{\rm IR} $   is an IR cutoff in \adst and $a$ is the radius of the boundary $S^2$. 
  In general, 
 a regularized volume of a  global AdS$_{p+1}$ space with  $S^p$ as its  boundary  is log  IR divergent  for   even $p$
 (discarding power divergences, see, e.g., \ci{Diaz:2007an}): \ \ 
$ \vol({\rm AdS}_{p+1})=
\frac{2(-\pi)^{p/2}}{\Gamma(1+{p}/{2})}\log \RR\,. $ } 
\be \la{91}
\F_{0} = \TT_2\ {\rm \vol}({\rm AdS}_3) = -2\pi \TT_2\  \log \RR  \ , \qquad {\rm \vol}({\rm AdS}_3) = - 2 \pi   \log \RR , \qquad 
 \te \TT_2=  {L^3  T_2} = {2\ov \pi} N   \ . 
\ee
Since \adst is a homogeneous   space, all quantum corrections  to $F$   will  also be  proportional to ${\rm \vol}({\rm AdS}_3)$, i.e. 
\ba
&F= F_0+ F_1 + F_2 + ... =  f(\TT_2)\,  {\rm \vol}({\rm AdS}_3) \equiv - \tfrac{1}{3} \bb  \, \log \RR  \ ,\la{02}\\ % \ \ \ \ \ \ \ \ 
&\ f(\TT_2) = \TT_2  f_0 + f_1 + (\TT_2)^{-1} f_2 + ...  \ , \ \ \ \ \ \ \ \te  \bb =  {6\pi} f= N\, \bb_0 + \bb_1 + N^{-1} \bb_2 + ... \ .  \la{04} \ea
Since $\RR$  plays the role of a  UV cutoff in the boundary theory, $\bb$   may be  interpreted as  an $S^2$ 
defect  conformal anomaly  in the (2,0) theory. 

Fixing the static gauge  and expanding the M2 action 
to quadratic order in fluctuations  near AdS$_3$   background 
one  finds  that the spectrum of the resulting  \adst  fields   consists of  4  
bosons  $x^i$ with $m_b^2=3$,   4 bosons $y^a$  with $m=0$     and
8 fermions  $\theta$ with  $m_f= {3\ov2}$.  This  spectrum of transverse fluctuations of the M2 brane subject to the standard Dirichlet b.c. 
 is  in direct
correspondence with a protected supermultiplet (that includes the displacement operator) of operator insertions on the  defect 
surface  \ci{Drukker:2020swu}.
The resulting 1-loop correction  $F_1$ in \rf{02}  is given by 
\be \la{92}
F_{1}=  \ha \Big[4  \log \det   ( - \nabla^2 + 3)   + 4  \log \det  ( - \nabla^2 ) -  8 \log \det \Delta_{1/2} \Big]  = f_1\,  {\rm \vol}({\rm AdS}_3)  \ . 
\ee
There  are  no  1-loop  log  divergences in 3d   so that $f_1$  is  finite  when  computed using the 
standard $\zeta$-function regularization.\foot{Using  heat-kernel cutoff, one finds that the leading cubic  divergence cancels out   due to supersymmetric balance of degrees of freedom. Linear divergence   does not   automatically  cancel 
but is absent
 in an analytic  regularization like the $\zeta$-function one. Similar linear   divergence  was regularized away  using $\zeta$-function in a different  1-loop M2 brane computation in \ci{Giombi:2023vzu}.} 
As a result, one finds  that  \ci{Drukker:2020swu}
\be  f_1 = -\te  {3\ov 2\pi} \ , \qquad {\rm i.e.} \ \ \ \    \bb_1=- 9 \ . \la{06}
\ee
Combining \rf{91}  and \rf{06} we get the first two terms in \rf{01},\rf{03} and \rf{002}. 

Finding  the  2-loop  correction to  the free energy \rf{02}  requires expanding the M2  brane   action \rf{31} 
near \adst  surface to quartic order in the fluctuation fields $(x^i, y^a, \theta)$.
%\foot{A similar
% 2-loop  computation  of the free energy  of the GS    string  with AdS$_2$   minimal surface in 
% AdS$_5 \times S^5$ that  is  expected to be  related to the  strong-coupling expansion 
% of the circular Wilson loop in $\N=4$ SYM theory  (cf. \ci{Giombi:2020mhz})  is  discussed  in \ci{Beccaria:2026ffm}.
%}
Remarkably, like  for the  M2  brane in static gauge in  flat  target space case   \ci{Seibold:2024oyr,Beccaria:2025xry}
(and also as for  the  GS string  expanded  near AdS$_2$    in \adsp\   \ci{Beccaria:2026ffm})  there  is 
a  natural $\kappa$-symmetry gauge  in which  there are no cubic couplings in   the action. The M2   Lagrangian
  has then  the following symbolic  form\foot{Here all indices   are contracted with the induced 
    \adst metric and derivatives are  3d covariant   so there is a manifest \adst   symmetry 
  with fermions $\theta$  treated effectively  as a set of 8  Majorana  3d fermions. We rescaled  the fluctuations  by $\sqrt{\TT_2}$.} 
 \ba  
   L = & (\del x)^2  + m^2_{b}  x^2 + (\del y)^2  +  \bar \theta ( \slashed \nabla + m_f) \theta  + \TT_2^{-1} \Big[ (\del x)^4  + x^2 (\del x)^2 + x^4  + (\del x)^2 (\del y)^2  \la{5} \\ 
 &  +   (\del y)^4 + y^2 (\del y)^2 + (\del x  \del x  + x^2) (\theta \nabla \theta + \theta^2 ) + ...
     + \theta  \theta \theta  \nabla \theta   +   \theta\nabla  \theta\theta  \nabla \theta + ...     \Big]  + \OO ( \TT_2^{-2}) \ . \no
   \ea
   As a result, 
the relevant 2-loop diagrams are just the double-bubble  ``OO'' ones, i.e. are 
  given by products  of  (derivatives of)  two  bosonic, one bosonic  and one fermionic    and     two fermionic  propagators in 
AdS$_3$ at coinciding points. 

Since  in 3d the  propagators have no log divergences,  any UV   divergences in $F_2$
can therefore only be power divergences, which are absent in an analytic regularization such as dimensional regularization
%the 2-loop  correction $F_2$    may then   have only power divergences
%and thus will be  finite  assuming one uses  an analytic regularization like  dimensional regularization  
where one 
replaces \adst with AdS$_{d+1}$ with $d=2-2\eps$. 
%in which $\delta(0)$ terms are set to 0.
Below we  will use  %(like   in the string theory case in \ci{Beccaria:2026ffm}) 
 its dimensional reduction version by treating  fermions (and related  Dirac matrices) 
not as $d+1$   but as 3-dimensional  ones.
We will find  that  then   the 2-loop coefficient in \rf{04} is given by 
\ba
(f_2)_{_{\rm dred}}= 12\,\frac{d-2}{d+1}\gs_x^{2}+12\, \frac{(d-2)(d+7)}{(d+1)^{2}}\,\gs_x\,\sss
+24\,\frac{(d-2)^{2}}{(d+1)^{2}}\, {\rm G}_\theta^2 \ , \la{11}
\ea
where $\gs_x$ and $\sss$ are the  coincident-point limits  of the  massive scalar ($m_b^2 = 3$) and massive fermion  ($m_f= {3\ov 2}$) propagators  respectively. These   are finite for $\eps={2-d\ov 2} \to 0$ 
\be\la{12}
 \gs_x = -\frac{1}{2\pi}+\OO(\eps) \ , \qquad \qquad %\frac{3-4\LDR}{8\pi}\, \eps+\cdots, 
\qquad  \sss = \frac{1}{2\pi}+\OO(\eps ) \ . %\frac{1+2\LDR}{4\pi}\, \eps+\cdots, \ \  \qquad \LDR = \log(\pi e^{\gamma_{\rm E}}) \ . 
\ee
Since  $f_2$ in \rf{11} is proportional  to $d-2$,   it thus vanishes  in the $\eps\to 0$ limit, i.e.\foot{The same vanishing result  for $f_2$ is found also 
in the straightforward dimensional regularization as   near  3d  $\gs_x$ and $\sss$ do not  have poles in $1\ov \eps$.}
\be  \la{09}  \te     (\bb_2)_{_{\rm dred}}=3\pi^2 (f_2)_{_{\rm dred}} = 0  \ . \ee %v4
%This contradicts  the prediction  in \rf{03}  which implies that $\bb_2$ in \rf{04}     should   be equal to  $- 3 $.
This matches the prediction in \rf{002}.

%v4
One may wonder if this  result  depends on  a choice of regularization.
% mismatch is due to a  ``wrong''  choice of regularization. 
In general,  %The  main issue is   which 
the   world-volume  UV regularization should be  the   one   which 
 is  consistent  with underlying symmetries of the problem and thus hopefully 
with the  suggested AdS/CFT  interpretation  of  the M2 brane free   energy as capturing the value of the surface 
defect  anomaly. The 
dimensional reduction regularization appears, in fact,  to be a natural choice as it   should 
 preserve  the world-volume supersymmetry 
 that is present  in the M2 brane action expanded near a  supersymmetric minimal surface: this 3d supersymmetry 
  is   a remnant  of  the target space  supersymmetry  after fixing  a $\kappa$-symmetry gauge 
 \ci{Bergshoeff:1987cm,Bergshoeff:1987qx,
 Kallosh:1997ky,Kallosh:1997sw}.
 Below   we will show that the 
 %Still,  it  could  be that some other analytic regularization  % (like $\zeta$-function one  used at 1-loop order) 
% is  a   more  adequate one.   Below we will  explore this option  and conclude that the same   
 vanishing result  \rf{09}  is found also   in the  same $\zeta$-function  regularization  that was used  at the 1-loop level in \ci{Drukker:2020swu}. 
  %This    strongly suggests   that the vanishing of $f_2$ is indeed a  robust  conclusion. 
 
\iffa 
One  may  also suspect   that a    reason for  the   disagreement between \rf{03}  and \rf{09} 
may  be related to the   fact  that   the  formally  non-renormalizable   BST   M2  brane action 
  should be supplemented by higher derivative  counterterms \ci{Beccaria:2025xry}: these 
  may  then 
   also   contribute  to the 2-loop  value of the free energy evaluated on the \adst surface. % (see comments in section 5). 
%One may also conjecture
 Alternatively,     there  may be some  profound reason   for   this  disagreement %why \rf{09} is in disagreement   with \rf{03} 
   that has to do with how  the   AdS/CFT   correspondence  is to be implemented  in the  context of  
   M2   brane partition functions     (cf. \cite{Gautason:2025per,Gautason:2025plx}). 
   \fi
   
   %A26
   %v4
 % Finally, the simplest  explanation  for how to reconcile our  vanishing 2-loop  result with  the general expression 
 % for the b-coefficient in \rf{001}   was suggested to us  by J. van Muiden \ci{vanMuiden:2026ffm}.

  \
  
The rest of this paper is organized as follows. 
In section 2  we will  find the 2-loop  contribution to free energy \rf{02} from the 4+4   transverse bosonic   fluctuations in the static gauge. 
The fermionic 2-loop contribution will be  computed  in section 3. The total  result for the 2-loop coefficient $f_2$ will be presented in section 4. 
In section 5 we will   make some concluding remarks.
% discussing  
%several possible  explanations   for the above disagreement.

In  appendix A   we shall  summarize the expressions for the  bosonic and fermionic Green's functions and their derivatives  in AdS$_{d+1}$ 
and specialize  them  to the case of   dimensional reduction  regularization in 
 $d=2-2\eps$   and the $\z$-function regularization in $d=2$. 
In appendix B  we will review  the  supercoset construction of the M2 brane action in \adsss 
 and  discuss its expansion  in powers  of fermions. 
In appendix C we will  summarize the  expressions  for the    quadratic and quartic fermionic correlators 
 at coincident points  that are used   in section 3.

\section{2-loop  contribution from bosons \la{sec2}} 

%%%%%%%%%%%%%%%%%%%%%%%%%%%%%%%%%%%%%%%%%%%%
\subsection{Expansion of  bosonic part of M2 brane action} 

Let us    recall   the expansion  of the bosonic   part of the M2 brane action  in   AdS$_7 \times S^4$ in 
the static gauge.  As in     \ci{Giombi:2017cqn,Drukker:2020swu}, 
for generality, let us 
consider a $p$-brane in AdS$_{D+1}\times S^n$  with  world volume  ending  along a $p$-dimensional   surface   at the boundary. 
%Following \cite{Giombi:2017cqn}  where the case of $p=1$ and $D=5$  was discussed, 
Let us choose the following  AdS$_{p+1}$-adapted  parametrization of AdS$_{D+1}$ (with radius 1) 
 \beg  \la{3399} 
ds^2_{D+1}  =\frac{ (1+\fo  x^2)^2}{(1-\fo x^2)^2} ds^2_{p+1}  + \frac{d x^i dx^i}{(1-\fo x^2)^2} \ ,
%\qquad ds^2_{p+1} =\frac{1}{\xxxx^2} (dx^2 + d\xx^v  d\xx^v)      \    , 
\eeg
where  %the indices of the  boundary coordinates  of  AdS$_{D+1}$  are split  into    $\mu=1, \dots$ and 
  $ i=1, \dots, D-p$.
In the static gauge where the  $p$-brane  world volume coordinates $\s^\m$ are identified with  the AdS$_{p+1}$ ones 
 the induced  metric is that of the AdS$_{p+1}$
%The minimal   surface  ending on  a $p$-plane  at the boundary  is 
\beg \la{34}  %\xx^v = x^v \ , \qquad  \xxxx= \xxx  \ ,   \qquad \xx^i=0\  ,  \qquad 
ds^2_{p+1}  \equiv  g_{\m\n}(\s)\, d \s^\m d\s^\n  =  ds^2_{\AdS_{p+1}} \ .
\eeg 
Then  the  volume  $S^{(1)}  $ part of  the  brane  action in \rf{31}  takes the form 
\beg \la{357} 
S^{(1)}   =  {\rm T}_p  \int d^{p+1} 
\s  \,  \sqrt{\det \Big[    \frac{(1+\fo  \xx^2)^{2}}{(1-\fo \xx^2)^{2}}  g_{\m\n}(\s)   + \frac{\del_\m \xx^i \del_\n \xx^i  }{(1-\fo \xx^2)^2}
+  
{  \partial_\m y^a\partial_\n y^a \ov (1+\tfrac{1}{4\rr^2}   y^2)^2}  \Big] } \equiv   % {\rm T}_p
 \int d^{p+1}x   \sqrt{ g}\  L  . 
\eeg
Here $y^a$   are coordinates of $S^n$  and $\rr$  is its radius in units of the radius  $L$ of AdS$_{D+1}$  (cf. \rf{452}). $L$  
 is absorbed into  the dimensionless effective tension $ {\rm T}_p = L^{p+1} T_p  $, cf. \rf{91}.

Expanding \rf{357}   in powers of the fluctuations  $\xx^i$ and $y^a$ we get 
\ci{Drukker:2020swu}\foot{Here $\m,\n$ indices are contracted by AdS$_{p+1}$ metric $g_{\m\n}$ and we rescaled fluctuations by 
the square root of the  tension.}
\begin{align}   L &=L_{2\rm b}  + {\TT}_p ^{-1} L_{\rm 4b}    + \ldots   \ , \qquad \qquad   L_{\rm 4b}  = L_{4\xx}   + L_{2\xx,2y} + L_{4y}  \ , \la{334} \\
L_{2\rm b}&=\tet  \frac{1}{2} \big[ \del^\m \xx^i \del_\m \xx^i  +  (p+1)\, \xx^i \xx^i\big] + \frac{1}{2}  \del^\m y^a \del_\m y^a\ ,  \la{36} 
\\ 
L_{4\xx} &=\tet  \frac{1}{8} (\del^\m \xx^i \del_\m \xx^i )^2  
           - \frac{1}{4}  (\del^\m \xx^i \del_\m \xx^j) \; (\del^\n \xx^i \del_\n  \xx^j)
\tet + \frac{1}{4}  p\,  \xx^i \xx^i  \, \del^\m \xx^j \del_\m \xx^j + \frac{1}{8} (p+1)^2   \xx^i \xx^i\, \xx^j \xx^j \ , 
\label{377}            
\\
L_{2\xx,2y}&=\tet  \frac{1}{4}  (\del^\m \xx^i \del_\m \xx^i )\,(\del^\n y^a \del_\n  y^a) 
    - \frac{1}{2}   (\del^\m \xx^i \del_\m y^a) \; (\del^\n \xx^i \del_\n  y^a)     
+  \tfrac{1}{4}   (p-1)   \xx^i \xx^i  \, \del^\m y^a \del_\m  y^a  \ , \la{3888}
\\
L_{4y} &= \tet
\frac{1}{8}   (\del^\m y^a \del_\m y^a)^2
-\frac{1}{4}  (\del^\m y^a \del_\m y^b) \; (\del^\n y^a \del_\n  y^b)-\frac{1}{4\rr^2} y^b y^b\ \del^\m y^a \del_\m y^a
\ . \la{39}
\end{align}
The  case of the AdS$_2$ string in AdS$_5 \times S^5$   considered in  \cite{Giombi:2017cqn,Beccaria:2026ffm}  corresponds to $p=1,\ D=4,  \ n=5, \ \rr=1$ 
while in  the present  case of  \adst  M2 brane in AdS$_7 \times S^4$   (cf. \rf{452}) 
\be  p=2\ , \qquad  \ \ \ D=6\ ,\ \ \ \  \qquad n=4   \ , \qquad \rr=\ha\ .   \la{2.1} \ee 
We thus   
  get  4 massive  transverse \adss  fluctuation fields  $\xx^i$  (with   $m^2=3$)  and 4 massless  $S^4$ fields  $y^a$
propagating in the induced \adst geometry.

The WZ term in \rf{3200}    may be written as  \ci{Drukker:2020swu} ($Y^m Y^m=1, \ \ m=1, ..., 5$)
\ba 
S_2= iT_2  \int  C_3  =   iT_2 \int  {\cal F}_4  & =
    \tfrac{i  }{ 64 } \TT_2 \int d^4 \s  \  \ep_{mnpqu}\, \ep^{\m\n\l\rho}\,   Y^m \del_\m Y^n \del_\n Y^p \del_\l Y^q \del_\rho Y^u \  \no \\
&\la{40} =   \tfrac{ i }{ 4 } \TT_2   \int d^3 \s \, \ep^{\m\n\l}\,  \ep_{abcd} \,  y^a \del_\m y^b  \del_\n y^c  \del_\l  y^d + \OO(y^5)   \ , 
\ea
where $ 
Y^5=  \frac{1-  y^2}{ 1 +  y^2}\ , \ \   
Y^a=  \frac{ 2y^a}{ 1+   y^2}  . 
$
The quartic term in \rf{40} (see also \rf{2024})
  will not contribute to the  2-loop  free energy $F_2$ in \rf{02}
%(given by bubble graphs) 
 due to the  symmetry of  the resulting contractions $\langle  y^a y^b \rangle \sim \delta^{ab}$. 
As a  result,   the  bosonic contribution  to $F_2$   will  come only from \rf{36}--\rf{39} and thus   will   have  the form   which 
is   universal  in $p$.

\subsection{Expectation value of  quartic  bosonic  terms }

Defining   the Euclidean  M2 brane  partition function as $Z= e^{-F} = \int [dx\,dy\,d\theta] \ e^{-S}$ where $S$ is  the action in \rf{31} 
  the 2-loop contribution $F_2$  to free energy in \rf{02}   may be written as   an  expectation 
   value of the  quartic term in the expansion of the   action
   \ba
   &S_4=\TT_2^{-1}\int d^3 \s\, \sqrt g \, L_4  \ , \qquad \qquad    L_4 =   L_{\rm 4b} +  L_{\rm 4bf} +   L_{\rm 4f} \ , \la{222}\\
& F_2 = \TT_2^{-1} \int d^3 \s\, \sqrt g \, \langle  L_4  \rangle = \TT_2^{-1}\, \vol(\rm AdS_3)\,  \langle  L_4  \rangle    \ , \ \ \ \ \ \ \ \  \ \ \ 
   f_2 = \langle L_4 \rangle   \ , \la{221}
\ea
where    we used   that since the M2 brane action  written in terms of the  \adst metric    has  constant coefficients
 and  that \adst is a homogeneous space the \adst  volume factor factorizes 
(cf. \rf{02},\rf{04}).   $L_{\rm 4b}$  in \rf{222}   stands  for the  quartic  bosonic term  in \rf{334}   while 
the terms involving fermions (cf. \rf{5})  will be discussed in section 3 below. 

Then from \rf{334}   we find   for the purely bosonic  contribution 
%Explicitly, we find the following  generalization  of \rf{d6}  (that corresponds to the case of $p=1$, $r=1$; here we use  Minkowski overall sign as in \rf{d6}, i.e. take minus  of \rf{36}) 
\begin{align} \label{999}
f_{2, 4\rm b} =  \langle {L}_{4\rm b} \rangle =&\te   {1\ov 8 } (p^2 -1) N^2_{x}\rhg_{x}^2  -    {1\ov 4 } (p+1)^2 N_{x}\rhg_{x}^2\no \\
  & \te +\frac{1}{4}p (p+1) N_{x}^2 \rG_{x}\rhg_{x}
  +\frac{1}{8} (p+1)^2 N_{x}(N_{x}+2) \rG_{x}^2  \\
 &\te + {1\ov 4 } (p^2-1)  N_{x} N_y \rhg_{x}\rhg_{y}
  + {1\ov 4 } (p^2-1)  N_{x} N_y \rG_{x}\rhg_{y}\no \\
&\te    +  {1\ov 8 } (p^2 -1) N^2_{y}\rhg_{y}^2  -    {1\ov 4 } (p+1)^2 N_{y}\rhg_{y}^2 -\frac{1}{4 r^2} (p+1) N_{y}^2 \rG_{y}\rhg_{y} \ .\no 
  \end{align}
Here  $N_x=D- p$ and $N_y=n$   are the numbers of the  corresponding fluctuations around AdS$_{p+1}$  in AdS$_{D+1} \times S^n$, i.e.  in the  M2  brane case  \rf{2.1} 
% AdS$_{D} \times S^{n}$  ($D=7, n=4$, or $D=4, n=7$ or $D=5, n=5$)  we  have for  AdS$_{p+1}$   brane  \be 
%N_x= D- p-1 \ , \ \ \ \   N_y= n \ ,  \la{d78} \ee
% i.e. in the  case of M2  brane  in $AdS_7 \times S^4$ 
\be\la{96}
p=2\, :  \qquad \ \ \ N_x=4, \ \qquad \ \   N_y=4  
\ . \ee 
$\rG_{x,y}$ and $\rhg_{x,y}$ are the coincident limits of the corresponding scalar Green's functions in AdS$_{p+1}$
%defines as 
\ba 
&\langle  x^i (\s)  x^j (\s') \rangle = \delta^{ij} G_x(\s, \s') \ ,   \qquad  G_x(\s,\s)  = \rG_x  \ , \no \\
&  \del_\mu \del'_\nu G_x (\s, \s') = g_{\m\n} \tilde G_x (\s, \s'), \qquad 
\tilde G_x (\s,\s)  = \rhg_x\ ,  %\qquad\qquad  \phi = (x, y)
  \la{100}
\ea
 and similarly for $\langle  y^a (\s)  y^b  (\s') \rangle = \delta^{ab} G_y (\s, \s')$, etc. 
 
Note that \rf{999}  written  for   general $p$ is not the same as  the result   found  using dimensional regularization near  a particular 
 value of $p$: in  the latter case  we  are first  to specify 
 $p$ (i.e. fix the coefficients in the  fluctuation Lagrangian in \rf{334}--\rf{39},  in particular the mass  of $x^i$ fluctuations as $m^2= p+1$) 
   and then   replace AdS$_{p+1}$  by  AdS$_{d+1}$  with   $d=p- 2 \eps$, i.e. 
     extend  the  indices of  derivatives and $g_{\m\n}$   to  $d+1$  values.   
 Then  $g_{\m\n} g^{\m\n}= d+1$  giving  a generalization  of \rf{999}
  \begin{align} \label{771}
 f_{2,4\rm b}  
  =&\te   {1\ov 8 } (d^2 -1) N^2_{x}\rhg_{x}^2  -   {1\ov 4 } (d+1)^2 N_{x}\rhg_{x}^2\no \\
  & \te +\frac{1}{4}p (d+1) N_{x}^2 \rG_{x}\rhg_{x}
  +\frac{1}{8} (p+1)^2 N_{x}(N_{x}+2) \rG_{x}^2  \\
 &\te + {1\ov 4 } (d^2-1)  N_{x} N_y \rhg_{x}\rhg_{y}
  +{1\ov 4 } (d+1) (p-1)  N_{x} N_y \rG_{x}\rhg_{y}\no \\
  &\te  +  {1\ov 8 } (d^2 -1) N^2_{y}\rhg_{y}^2  -    {1\ov 4 } (d+1)^2 N_{y}\rhg_{y}^2\
    -\frac{1}{4 r^2} (d+1) N_{y}^2 \rG_{y}\rhg_{y} \ .\no 
  \end{align}
  %Specifying to $p=2, \ \rr=\ha $ and
   Introducing  the notation for the ``equation of motion'' $(-\nabla^2 + m^2 ) G\big|_{\s=\s'}$   or ``$\delta(0)$''  combinations    
  \be \la{88}
   \rh_{x} \equiv   \rhg_x  + {p+1\ov d+1}  \rG_{x} \ , \qquad \qquad  \rh_{y} \equiv   \rhg_y  \ , \ \ \ \   \ee
  we may  represent \rf{771} as 
  \begin{align} % \label{77}
 f_{2,4\rm b}  
  =\te   {(d-p) (p+1) \ov 4 (d+1) } N^2_x\,  \rG_x^2   &\te  +  \rh_x \Big\{ \big[ {1\ov 8 } (d^2 -1) N^2_{x} -   {1\ov 4 } (d+1)^2 N_{x} \big]      \rh_x         \no \\ &\te   
 \qquad   +    
  \big[ {1\ov 4} (  2p - d +1) N^2_x  + \ha  (d+1) (p+1) N_x                \big] \rG_x \Big\}   \no \\
  &\te  + \rh_y \Big\{ {1\ov 4 } (d^2-1)  N_{x} N_y    \rh_x   + \ha (p-d)  N_x N_y \rG_x  
  \no \\
  &\te  \qquad + \big[ {1\ov 8 } (d^2 -1) N^2_{y} -    {1\ov 4 } (d+1)^2 N_{y}\big] \rh_{y}
    -  {1\ov 4 \rr^2}(d+1) N_{y}^2 \rG_{y}\Big\}  \ .\la{77}
  \end{align}
Thus the  total expression for any $p$ is given  by the   sum of the first $\rG_x^2 $  term  with 
 the two extra   contributions proportional to $\rh_x$ and $\rh_y$, i.e. to the  ``$\delta(0)$''   terms.  

%For example, for the AdS$_5 \times S^5$  case  where  $p=1, \ d= 1-2\eps$  and $\rh_{x} = - {2\ov d+1}  \rG_{x}, \rh_y=0$
%we reproduce the expression in dimensional regularization in \rf{449}. 
Using that in the dimensional regularization where $d=p-2\eps$  (see appendix \ref{apG0})
\be   \rh_x=0 \ , \ \ \ \qquad    \rh_y=0 \ , \la{333} \ee
%$-\nabla^2 G_x + m^2 G_x=0, \ \  -\nabla^2 G_x =0$  we then get 
  % for  $m^2= p+1$   and    $d= p -2 \eps$   that 
   we  conclude that \rf{77} reduces to 
\be \la{809}
%  \langle \mathcal{L}_{4\rm b} \rangle 
  (f_{2,4\rm b} )_{_{\rm dreg} } = {(d-p) (p+1) \ov 4 (d+1) } N^2_x\,  \rG_x^2  = - {\eps\ (p+1) \ov 2 (p+1 - 2 \eps) } N^2_x\,  \rG_x^2
  \ . \ee %v4
  For $p=2$, i.e.  in $d+1=3-2\eps$  dimensions,  $\rG_x$ has no pole in $\eps$  (see \rf{12}  and appendix  \ref{apG}).  
  We  thus 
   conclude that \rf{809} vanishes   in the limit $d\to p$. The same  conclusion applies
   to other $p>1  $ brane cases.\foot{Note that a similar   general  expressions \rf{77},\rf{809} 
    is   found also  in the  case of the 
  AdS$_2$   string  in   AdS$_5 \times S^5$ 
    \ci{Beccaria:2026ffm}   but there $\rG_x$ contains a pole and thus  the analog of \rf{809} 
   is UV divergent.}
  
   The expressions  for the  fermion  contributions discussed below  will  have  similar  form: % the same pattern:
   % as in the string case, i.e.  
   they  will also be  proportional to $d-p= -2 \eps$  and thus  will  vanish for $d\to p=2$  (cf. \rf{11}). 
 % that will mean that the  total 2-loop  contribution is zero. 
  
 %May be like  in string case   the  expectation  value of this term reduces to $(\del x \del x + (p+1) x^2)  \bar \theta G_* \theta $  so 
  %vanishes on-shell  and thus gives 0 in dim reg. 

%%%%%%%%%%%%%%%%%%%%%%%%%%%%%%%%%%%%%%%%
\section{2-loop  contribution from fermions \la{sec3}} 

\subsection{Expansion of the fermionic part of the M2 brane action \la{sec31}}

The   explicit form  of the  M2 brane    action 
\cite{Bergshoeff:1987cm,Bergshoeff:1987qx}  in \adsss
may be found following  \ci{deWit:1998yu,Claus:1998fh}
(see also \ci{Sakaguchi:2003hk,Sakaguchi:2004bu}). 
We  review its structure  in appendix  \ref{apM}  and discuss   the expansion to  quartic order in the fermion field $\theta$. 

Let us  first  introduce  the notation.  Starting with the 
11d Majorana spinor $\theta$   we will analytically continue to the Euclidean signature, i.e.  consider 
the   $32\times32$ Dirac matrices that  satisfy the Clifford algebra  ($A  = 0, \dots, 10 $)
 % {\bf rename  primes to nonprimes} 
\iffa 
\footnote{Note that in  the notation 
where $\theta$ is a product of 7d and 4d spinors one has 
$ \Gamma_{a} = \gamma_{5} \otimes \gamma_{a}   , \quad \Gamma_{a} = \gamma_{a}\otimes \id  , \quad \gamma_{5} = \gamma_{7}\dots\gamma_{10}$, $
     C = C\otimes c \ , \quad \gamma_{a} = -C^{-1}\gamma_{a}^{T}C \ , \quad \gamma_{a} = -c^{-1}\gamma_{a}^{T}c \ ,
    $
%\end{align}
where $\gamma_{a}$ are $8\times 8$ Dirac matrices of 7d Clifford algebra, $\gamma_{a}$ are $4\times 4$ Dirac matrices of 4d Clifford algebra,  and  $C$ and $c$ are antisymmetric and symmetric charge conjugation matrices respectively.}
\fi 
\begin{align}
    & \{ \Gamma_{A} , \Gamma_{B} \} = 2\delta_{AB} \ , \qquad
     \Gamma _{A} =  \big(\Gamma_{\halpha}, \Gamma_{i}, \Gamma_{a}\big)  \ , \quad \halpha =0, 1, 2 \ , \quad i = 3, \dots,6 \ , \quad a = 7, \dots, 10 
    %\Gamma_{A} = \{\Gamma_{a}, \Gamma_{a}\} \ , \qquad a = 0, \dots 6 \ , \qquad a = 7, \dots, 10 
      \ .\la{32}
\end{align}
 To 
  account for the Wick rotation, we  assume that the  charge conjugation matrix  $C$ satisfies the same properties as in the Lorentzian case:
$ C^{T} = - C , \ \ \Gamma_{A} = - {C}^{-1}\Gamma_{A}^{T} C \ .
$
Let us define  the matrices  $  \Gst$   and $\Gamma$  which satisfy  
\begin{align}
    & \Gst \equiv  \Gamma_{7}\G_8 \G_9\Gamma_{10}  \ , \quad 
    \Gst^2 = \id \ , \quad  \Gst =  C^{-1}\Gst^{T} C \ , \quad      [\Gst,\Gamma_{\halpha} ]=  [\Gst,\Gamma_{i} ]= 0 \  ,   \quad \{\Gst,  \Gamma_{a} \}=0
 \ , \la{35}\\ 
%\end{align}
%We also note that $\Gamma_{*}$ on a Clifford subalgebra formed by $\Gamma_{a}$ acts as a Hodge star operator. Namely:
%\begin{align}
 &  \qquad  \Gamma_{a}\Gamma_{b}\Gamma_{c}\Gamma_{d} = \epsilon_{abcd}\Gamma_{*} \ ,\ \ \  \qquad \Gamma^{a}\Gamma_{*} = \tfrac{1}{3!}\epsilon^{abcd}\Gamma_{bcd} \ ,   \no 
\\ 
\la{33} 
  &  \Gamma \equiv  i \Gamma_{0}\Gamma_{1}\Gamma_{2} \ , \qquad \Gamma^2 = \id \ , \qquad
   \Gamma =  C^{-1}\Gamma^{T}  C \ ,\\
%\end{align}
%which is further used for the $\kappa$-symmetry projector in the GS action. 
%We will   use  also   the following  3+4+4 subsets  of the 11d  Dirac matrices 
%\begin{align}
\no %\la{36}
% & \la{36}   \Gamma _{A} =  \big(\Gamma_{\halpha}, \Gamma_{i}, \Gamma_{a}\big)  \ , \qquad \halpha =0, 1, 2 \ , \qquad i = 3, \dots,6 \ , \qquad a = 7, \dots, 10 \ ,
%\\ \no 
    &[\Gamma,  \Gamma_{\halpha}]=0    \ , \quad \{\Gamma,  \Gamma_{i}\} =  \{\Gamma,  \Gamma_{a} \} =0 
  \ , \quad [ \Gamma, \Gamma_{*}] = 0  , \qquad 
    \te  \Gamma^{\halpha}\Gamma = \frac{i}{2!}\epsilon^{\halpha\hbeta\hgamma}\Gamma_{\hbeta\hgamma}  \ , \quad \epsilon^{012} = 1 \ .
\end{align}
We shall choose the following   $\kappa$-symmetry  gauge % in the M2  brane action
 that complements  the bosonic 
 static gauge  in    the same way as   in  the   flat  target space  case in  \ci{Seibold:2024oyr,Beccaria:2025xry}  %\ci{Seibold:20240yr,Beccaria:   } 
 % is imposed by requiring the fermionic fluctuations $\theta$ to satisfy:
\begin{align} \label{214}
    \Pc\,  \theta = 0 \ , \qquad \Pc \equiv  \tfrac{1}{2}\big{(}\id + \Gamma\big{)} \ , \qquad \Pc^2 =\Pc \ . 
\end{align}
Remarkably, like in the  flat space case, the resulting 
 expansion of the M2  brane action  (see appendix \ref{apM1})
 will then  have    no cubic boson-fermion-fermion coupling terms. This   substantially simplifies  the computation of the corresponding 
  2-loop  free energy. 
  
The quadratic   fermionic term in the action can be expressed in terms of the 
   analog of the  massive \adst  Dirac operator  with  $m_f= {3\ov 2}$  \ci{Forste:1999yj,Drukker:2020swu} defined  by  the following 
    generalized covariant derivative\foot{One can  formally view  the 
 16-component   fermion $\theta$ (remaining   after the gauge fixing \rf{214}) as a 
  collection of  8  2-component    3d Majorana   spinors.}
\be
 \mk D_{\alpha} = \nabla_{\alpha}+\tfrac{1}{2}\G_{*}\G_{\alpha},\qquad  \G_\a \equiv e^{\halpha}_\alpha \G_\halpha \ , \qquad  \qquad \slashed{\mk D} = \slashed{\nabla}+\tfrac{3}{2}\G_{*} \ . \la{37}
\ee
 Here   $e^\halpha_\alpha$ is the  \adst   3-bein ($g_{\a\b} = e^\halpha_\alpha e^\hbeta_\b \delta_{\halpha \hbeta}$) 
   and   $\nabla_\a$ is the \adst    spinor covariant derivative.
   Defining 
\ba    &  h_{\alpha\beta}^{(2b)} = h_{\alpha\beta}^{(2x)}+h_{\alpha\beta}^{(2y)},  \qquad \qquad  \la{41}
\qquad h_{\alpha\beta}^{(2\theta)} = \bar\theta\G_{\alpha}\mk D_{\beta}\theta, \\
& 
h_{\alpha\beta}^{(2x)} = \partial_{\alpha}{x}^i  \partial_{\beta}{x}^i +{x}^{2}g_{\alpha\beta}, \qquad \la{42} 
\qquad h_{\alpha\beta}^{(2y)} = \partial_{\alpha}{y}^a \partial_{\beta}{y}^a, 
%\mc M_{+}^{2} &= \G_{*}\G_{\alpha}\theta\bar\theta\G^{\alpha}-\frac{1}{2}(\G_{\alpha\beta}\theta\bar\theta\G^{\alpha\beta}+\G_{ij}\theta\bar\theta \G^{ij}-\frac{1}{\rr}\G_{ab}\theta\bar\theta\G^{ab})\G_{*},
\ea
one   finds (see \rf{473}--\rf{435})  that the 
   quadratic and quartic terms   in the expansion of the  M2 brane Lagrangian 
 that supplement   the bosonic terms in  \rf{334}   can be represented as (here we specify to the  M2 brane case of $p=2, \ \rr= \ha$) 
\ba
  L &=L_{2}  + {\TT}_2 ^{-1} \big(  L_{\rm 4b}  + L_{\rm 2b, 2f}  + L_{\rm 4f}\big) + ...  \ , \la{43}\qquad \qquad 
    L_{2} = \tfrac{1}{2}g^{\alpha\beta}(h_{\alpha\beta}^{(2b)}+h_{\alpha\beta}^{(2\theta)}),
\\
 L_{\rm 2b, 2f} &=\te  \frac{1}{8}g^{\alpha\beta}\big(h^{(2x)}_{\alpha\beta}- 2 h_{\alpha\beta}^{(2y)}\big)\bar\theta\G_{*}\theta
+\frac{3}{8}{x}^{2}g^{\alpha\beta}h_{\alpha\beta}^{(2\theta)}
+\frac{1}{8}g^{\alpha\beta}g^{\gamma\delta}h^{(2b)}_{\alpha\beta}\, h_{\gamma\delta}^{(2\theta)}
-\frac{1}{4}h^{(2b)\alpha\beta}h_{\alpha\beta}^{(2\theta)} +...\ ,\la{44} \\
 L_{\rm 4f} &= \te \frac{1}{96}g^{\alpha\beta}\bar\theta\G_{\alpha}\mc M_{+}^{2}\mk D_{\beta}\theta-\frac{1}{16}g^{\alpha\delta}g^{\beta\gamma}h^{(2\theta)}_{\alpha\beta}h^{(2\theta)}_{\gamma\delta}
+\frac{1}{16}(g^{\alpha\beta}h_{\alpha\beta}^{(2\theta)})^{2}\ .  \la{45}
\ea
We   defined    the  following  fermionic matrix (see \rf{b20}) 
\be\la{38} 
\mc M_{+}^{2} = \G_{*}\G_{\alpha}\theta\ \bar\theta\G^{\alpha}-\tfrac{1}{2}(\G_{\alpha\beta}\theta\ \bar\theta\G^{\alpha\beta}+\G_{ij}\theta\ \bar\theta \G^{ij}-  2 \G_{ab}\theta\ \bar\theta\G^{ab})\G_{*}, \qquad [\Gamma, \mc M_{+}^{2}]=0 \ . 
\ee
In \rf{44}  dots stand for terms in \rf{2025}  that  will not contribute to   the   2-loop free energy (they vanish upon use of $\langle  x^i x^j \rangle \sim \delta^{ij}$,  $\langle  y^a  y^b \rangle \sim \delta^{ab}$, 
$\langle  x^i  y^a \rangle =0 $)
 so we   will   omit them in what follows.

Explicitly, we  get from \rf{44},\rf{45}
\ba
 L_{\rm 2b, 2f} 
 %%%%%%%%%%%
= &\te  \frac{1}{8}(\partial_{\a}x^i  \partial^{\a} x^i +3{x}^{2})\,(\bar\theta\G_{*}\theta+\bar\theta\slashed{\mk D}\theta)
+\frac{1}{8}\partial_{\a}{y}^a\partial^{\a}{y^a}(\bar\theta\slashed{\mk D}\theta- 2 \bar\theta\G_{*}\theta)
\lp
%%%%%%%%%%%
+\te \frac{1}{8}{x}^{2}\bar\theta\slashed{\mk D}\theta
-\frac{1}{4}(\partial^{\alpha}{x^i}\partial^{\beta}{x^i}+\partial^{\alpha}{y^a}\partial^{\beta}{y^a})\, \bar\theta\G_{\alpha}\mk D_{\beta}\theta\ ,  \la{46}
\\
 L_{\rm 4f} 
= &\te \frac{1}{96}\bar\theta\G^{\alpha}\G_{\beta}\G_{*}\theta\ \bar\theta\G^{\beta}\mk D_{\alpha}\theta
-\frac{1}{192}\bar\theta\G^{\alpha}\G_{\beta\gamma}\theta\ \bar\theta\G^{\beta\gamma}\G_{*}\mk D_{\alpha}\theta
-\frac{1}{192}\bar\theta\G^{\alpha}\G_{ij}\theta\ \bar\theta \G^{ij}\G_{*}\mk D_{\alpha}\theta\lp
\te +\frac{1}{96}\bar\theta\G^{\alpha}\G_{ab}\theta\ \bar\theta\G^{ab}\G_{*}\mk D_{\alpha}\theta
-\frac{1}{16}\bar\theta\G^{\alpha}\mk D^{\beta}\theta\ \bar\theta\G_{\beta}\mk D_{\alpha}\theta +\frac{1}{16}(\bar\theta\slashed{\mk D}\theta)^{2} \ . \la{47}
\ea
Anticipating  the  use of dimensional reduction regularization  here   the indices  are   contracted  using  $g_{\alpha\beta}$  of AdS$_{d+1}$  with $d=2-2\eps$
unless they 
are  contracted with  $\G_{\alpha}$  matrices that restrict them to AdS$_3$  (Dirac algebra for $\G_\a$  is  assumed to  be done in 3d).

\subsection{Expectation values of the  fermionic   terms \la{sec32}}

To find the 2-loop contributions  of \rf{46} and \rf{47}  let us  define the basic  fermionic correlators
at coincident points 
  that complement the bosonic ones in \rf{100},\rf{88} 
\ba
\la{610}
& \langle\theta\, \bar\theta\rangle = -\G_{*}\sss, \qquad \langle \mk D_{\alpha}\theta\, \bar\theta\rangle = -\G_{\alpha}\ssss ,\qquad
 \langle\mk D_{\alpha}\theta\,  \mk D_{\beta}\theta\rangle = \big(\tgt\, \G_{\alpha\beta}+\bgt\, g_{\alpha\beta}\big)\, \G_{*}\, C^{-1} \ , 
\ea
where   factors  of gauge-fixing projector  \rf{214}   are implicit. 
In dimensional reduction regularization 
 (see  appendices \ref{apG1} and \ref{apF})
\ba
\la{611}\te
%\text{DRed}:\qquad \sss &= \sss, \quad\ \ 
 \ssss = \frac{d-2}{2(d+1)}\sss, \qquad\qquad  \ 
\tgt = \ssss \ ,  \qquad \qquad %\frac{1}{2}\frac{d-2}{d+1}\, \sss, \qquad\ \
  \bgt =  - {\ha} (d+2) \ssss   \ . % -\frac{1}{4}\frac{d^{2}-4}{d+1}\,\sss.
\ea
Note that these   vanish in the strict $d=2$ limit. 

As a result, we  get  from \rf{46}
\ba
\la{613}
\langle L_{\rm 2b,2f}\rangle 
= &\te \frac{1}{8}(d+1)  N_{x}N_{\theta} \rh_{x}\, (3\ssss+\sss)
+\frac{1}{8}N_{y}N_{\theta} (d+1) \rh_{y}\, (3 \ssss-  2 \sss)\lp\te 
%%%%%%%%%
+\big[ \frac{3}{8} + {9\ov 4 (d+1)} \big] N_{x}N_{\theta}\gs_{x}  \ssss
-\frac{3}{4}N_{\theta}(N_{x} \rh_{x}+N_{y}\rh_{y})\,  \ssss \ .
\ea
Similarly, the expectation value of \rf{47}  is found  not to depend on $\bgt$   and   is given by   (see   appendix \ref{apF}) 
\be
\la{617}
\langle L_{\rm 4f}\rangle =\te  \frac{1}{4}N_{\theta} (\ssss-3\tgt)\, \sss\, +\frac{1}{32}N_{\theta}^{2}\, \ssss (12\ssss+\sss)\, .
\ee

\section{Total  2-loop    contribution \la{sec4}}

Combining \rf{613} and \rf{617}  with  the bosonic contribution in \rf{77} (setting there $p=2, \ \rr=\ha$) 
we get the total  result for  the 2-loop coefficient $f_2$ in \rf{04} in a general regularization. 

In dimensional reduction regularization  where $d=2-2\eps$  we get from  \rf{613} and \rf{617}   (see \rf{88},\rf{611}; cf. \rf{77},\rf{809})  
\ba
\la{614}
%[(f_{2})_{ \rm 2b,2f} ]_{\rm dred} =
 \langle L_{\rm 2b,2f}\rangle_{_{\rm dred}} = &  \frac{3(d-2)(d+7)}{16(d+1)^{2}}\,N_{x}N_{\theta}\,\gs_x \sss\  , \\
\la{618}
\langle  L_{4f}\rangle_{_{\text{dred}}} = & -\frac{(d-2)}{4(d+1)}N_{\theta} \Big[1- \frac{7d-11}{16(d+1)}N_{\theta}\Big]\, \rG_\theta^{2}.
\ea
where \be   N_x=N_y=4 \ , \qquad \ N_\theta =16 \ . \la{666} \ee
Like  the  bosonic contribution  \rf{809}  the  fermionic contributions to $f_2$  in \rf{614} and \rf{618}  are all proportional to $d-2$. 
Summing up \rf{809},\rf{614},\rf{618}  and using \rf{666}   we   find that  the total  2-loop contribution in dimensional reduction regularization 
 is   given by\foot{Similar  expression  is  found  in  the standard  dimensional regularization where  instead of  $ \ssss = \frac{d-2}{2(d+1)}\sss$ in \rf{611}
 we have $ \ssss =0$.}
\ba
(f_2)_{_{\rm dred}}=  12\,\frac{d-2}{d+1}\gs_x^{2}+12\, \frac{(d-2)(d+7)}{(d+1)^{2}}\,\gs_x \sss
+24\,\frac{(d-2)^{2}}{(d+1)^{2}}\, \rG_\theta^2 \ .  \la{111}
\ea
It    thus   vanishes in the $d\to 2$ limit as $\gs_x$  and $\sss$ do not have poles   near   $d=2$  (see \rf{12}  and appendix \ref{apG})
\ba \te \no
{\rm  dred}: \qquad \ \  &\te \gs_x = - {1\ov 2\pi} + \OO(\eps)    \ , \qquad   \rh_x= 0 
  \ , \qquad   \gs_y = - {1\ov 4\pi} +  \OO(\eps)   \ , \qquad   \rh_y=  0 \ ,    \\
  &\te  \gs_\theta =  {1\ov 2 \pi}    +  \OO(\eps) 
\ , \qquad \ \ \  \rh_\theta =\tgt=   \OO(\eps)  \ . \la{112} \ea
The same   conclusion is reached also  in the $\z$-function regularization  (see appendix \ref{apG2}) where 
the entries  in \rf{77},\rf{613},\rf{617}   are %  to be computed with  (see  ....)\foot{Note that the values of $\gs_x$ and $\gs_\theta$
 the same as  in the dimensional regularization  for $d=2$  %(see  appendix \ref{apG})
\be \la{2223}\te 
{\rm \zeta-reg}: \qquad \gs_x = - {1\ov 2\pi} \ , \ \ \ \  \rh_x= 0 \ , \ \ \ \ \gs_y = - {1\ov 4\pi} \ , \ \ \ \  \rh_y= 0 \ , \ \ \ \ 
\gs_\theta =  {1\ov 2 \pi} \ , \ \ \ \  \rh_\theta =\tgt=0 \ , \ee
so that 
\be   
(f_2)_{_{\rm \zeta-reg}} =(f_2)_{_{\rm dred}}= 0 \ . \la{223}  \ee
%Let make few comments. 
Note that the   three terms  in \rf{111}   vanish separately in $d=2$.  
The vanishing of the bosonic contribution in \rf{77},\rf{809} 
appears to be a consequence of the special structure of the  action in \rf{357}--\rf{39} 
 leading to  the result  expressible in terms of the ``$\delta(0)$''  constants  
 $\rh_x,\rh_y$ % \rh_\theta$    
 (cf. \rf{100},\rf{88}) %,\rf{611}) 
% \be \la{919}
% (d+1) \rh_x +  3 \rG_x \equiv \delta_x, \qquad \ \ (d+1) \rh_y \equiv \delta_y 
% \ee
   that   vanish  for $d=2$ in   both  dimensional and $\z$-function regularizations. 
   Then a  similar  vanishing of the mixed  boson-fermion and  the fermion  contributions in \rf{111} may be attributed to  the supersymmetry of the M2 brane action.
 %In the fermionic case $\rh_\theta ={1\ov 3}  i \delta_\theta$ (cf. \rf{37},\rf{610})   also vanishes   in both regularizations. 
 
 Let us note also   that   contribution of a priori possible  ultralocal measure in the M2  brane  path integral  vanishes 
  as  ``$\delta(0)$'' terms   %(i.e. $\delta_x, \delta_y, \delta_\theta$)  
   are zero  in the above regularizations. 
   The  measure may need to be accounted  for  in a generic cutoff regularization like the  heat kernel one
   in which the  2-loop  correction to free  energy  may  contain  power divergences 
   %v2
   (see  appendix \ref{apG2}).\footnote{One might worry about additional 2-loop contributions to the free energy arising from 
   possible 1-loop finite counterterms introduced to impose renormalization conditions on the  
   2-point functions $\langle x^{i}x^{j}\rangle$
    and $\langle\theta\,\bar\theta\rangle$. A natural 
   requirement  is that the values of the masses  
   remain  at their tree-level values, so as to keep  the conformal dimensions
   of the fields dual to the  transverse fluctuations and fermions equal  to their  original values. 
   These fluctuations 
      belong to the  protected multiplet of operator insertions on the surface defect in the $(2,0)$ theory
   which includes   the displacement operator.  It turns out, in fact, 
   that 
    the resulting   counterterm   contributions carry  the 
    explicit  $d-2$ factors  as in (\ref{111})
     and therefore vanish
   for $d=2$.
   }

%%%%%%%%%%%%%%%%%%%
\section{Concluding remarks \la{sec5}}
%%%%%%%%%%%%%%%%%%%%%%%%%%%%
To summarize,  in this  paper we  computed the  2-loop   correction to the free energy  of the M2 brane in \adsss
 expanded near \adst minimal surface. 
 Despite  general   non-renormalizability of the M2 brane  theory defined by the 
 BST action   (as demonstrated at 2 loops  via 
   flat-space  S-matrix computation in \ci{Beccaria:2025xry})
 we  observed  that  the 2-loop $\AdS_3$ M2  free energy 
     is free from logarithmic  UV divergences and, in fact, vanishes in both  dimensional or $\zeta$-function regularization. 

Our result  implies    the vanishing of the $1/N$ 
 correction  to   the M2  brane  free  energy. This  agrees   with  the expectation that 
 it should  capture  the defect anomaly coefficient b   in the $U(N)$   \rf{002}  rather than
  $SU(N)$ \rf{03}  (2,0) theory.\foot{Same  should apply  to corrections to 2-point correlation functions of defect operators in (2,0) theory      discussed in 
\ci{Rigatos:2025sar,Chen:2024orp,Chen:2023yvw,Meneghelli:2022gps}.}

 % appears to be in  disagreement with   the naive \ci{Estes:2018tnu,Jensen:2018rxu,Chalabi:2020iie,Wang:2020xkc}
% value   \rf{03} of the  defect anomaly coefficient b   in the $SU(N)$ (2,0) theory
% but in fact  consistent with 

%   This   disagreement 
 % One   may  also worry  that probe  approximation 
%  may break down at high orders in $1/N$   requiring a modification of  background geometry  but such   a modification should  in fact apply only if the number of M2 branes is as large as $N$  \ci{DHoker:2008lup}.}
  %solution of that puzzle. 
 \iffa 
 One may suspect    that this disagreement  is   indicating  that the  ``defect AdS/CFT''   \ci{Giombi:2017cqn}
 or    quantum M2  brane 
 probe  description of the  boundary defect CFT as proposed in 
 \ci{Drukker:2020swu}  may be  breaking  down at higher  orders  in the semiclassical 
 M2 brane 
 expansion.\foot{One   may also   wonder if $N$ of the  M-theory 
  background    may get shifted  like  that  happens in the AdS$_4 \times S^7/\mathbb Z_k$ \ci{Bergman:2009zh}. However, 
  such  shift is not expected in the maximally supersymmetric \adsss   case.} 
 If true  that would  also  imply  that one  will   fail  to   reproduce     subleading  terms in the 
  boundary defect  correlators
  using perturbation theory  in   the \adst   M2 brane  world volume theory.\foot{Corrections to 2-point correlation functions of defect operators in (2,0) theory   were   discussed in 
\ci{Rigatos:2025sar,Chen:2024orp,Chen:2023yvw,Meneghelli:2022gps}.}  
 \fi 
 
 \iffa 
 One  reason  could be  that  since   the  M2 brane  theory based on BST  action   is not UV finite  starting  from   {2 loops}
 when expanded near a generic world-volume  background 
 %already in the   flat target space case
   \ci{Beccaria:2025xry}
   it may require  to be supplemented by  certain  higher-derivative counterterms.
 A  2-loop  counterterm (proportional to $\TT_2^{-1}$)   evaluated on the \adst  background  
  may then   produce a non-zero $1/N$   
 correction to the  free energy. 
  Unfortunately,  there is no known principle  (like integrability in 2d string case) 
 that   could   fix the  structure of  such   counterterms.
 % Given that  these  counterterms  may  also contribute  at higher loop  orders it would  then be 
  %hard to explain why  corrections  to the b-coefficient   truncate  at the 2-loop order as  implied by  \rf{03}. 
 \fi 
 
\iffa
An alternative  suggestion  is  that the standard  rules of the AdS/CFT correspondence  may need a modification  in this case.
% in the case of the M2 brane partition function 
There is a recent proposal in \cite{Gautason:2025per,Gautason:2025plx}  implying  that  comparison 
 between  gauge theory  localization results and the M2 brane  partition function (in particular, in the ABJM theory context) 
   should  take  into account its ensemble  interpretation. 
 \fi
 
\iffa  As was already mentioned  in the Introduction,  the simplest   possibility     is to interpret \ci{vanMuiden:2026ffm}
 our result   as    indicating  that 
  the quantum  M2  brane  probe   corresponds  to a surface defect in the  $U(N)$ rather than $SU(N)$ 
   boundary (2,0)  theory  where the b-coefficient      should  be given   by \rf{002}, i.e. should have no $1/N$  correction.
  \fi
  
% A further  insight  into this issue may  come from 
  It would be interesting to  perform   similar 2-loop computations in 
 the case of   M2 brane  wrapped on  either AdS$_2 \times S^1$   \ci{Giombi:2023vzu}  or $S^3/\ZZ_k$ \ci{Beccaria:2023ujc} 
  in the AdS$_4 \times S^7/\ZZ_k$  background  defining  the M-theory  dual  of   the ABJM theory. 
  An analogy  with  the  case discussed above  suggests   that in  the  static   gauge the expansion of the M2 brane 
   action  will   also   have   no cubic couplings and will be free of log UV  divergences. 
   Whether their sum vanishes in an analytic regularization remains to be established.\foot{In particular, 
   % and as a result  the 2-loop M2    free energy  may again vanish in an  analytic  regularization.
  for the $k=1$ case,  the   computation of the free energy  of  an  ``instanton'' M2 brane  wrapped on $S^3$  in AdS$_4 \times  S^7$ 
    may be   closely  related to the one in the present  paper   by a  certain analytic continuation (possibly modulo  zero and negative modes
     due to instability of  M2 on $S^3\subset S^7$ 
   \ci{Beccaria:2023ujc})     and thus the result 
    may  again vanish.}
  That  may then be consistent  with the ensemble conjecture in \cite{Gautason:2025per,Gautason:2025plx}.

 Another example    could be 
    the 2-loop   correction  to the  free energy of  an M2   brane  wrapped on  $S^1\times S^2$ in \adsss 
    where   the expression for the (2,0)    superconformal index   suggests that it   should    vanish \ci{Beccaria:2023sph}.
  The   methods  of the present paper should generalize  straightforwardly    to  these cases.

  % latitude? cusp ?  extra parameter ? 
% other  related observables 
 % to modifications  also beyond 1/N term   and so   one will still not
%match predicted b.
%Also, if there is a modification, why it is not visible at 1-loop order already?
% and 1-loop order is ok for sure multiplet of fluctuations is
%controlled by susy.
%In fact, I do not know examples  where there is  just tree+1-loop+2-loop correction  and no higher orders --  but there are many where  there is just tree+1-loop in susy setting. 
%[well, exception is eq 7.4 in https://arxiv.org/pdf/2303.16305   where there are corrections  to D^n R^4 terms from particular string loop orders; also in some N=2 susy + matter some non-holomorphic quantities  may have only up to 2-loop corrections]
%\ci{Drukker:2020atp}
%%%%%%%%%%%%%%%%%%%%%%%%%%%%%%%%%%%%%%%%%%%

\section*{Acknowledgements}
%v2
We  thank   N. Drukker, S. Giombi  and  J. van Muiden   for  useful  discussions.
We are  grateful   to  J. van Muiden  for  comments on the draft   and an  important  suggestion. 
 MB is supported by the INFN grant GAST. 
 SAK acknowledges the support of the President's PhD Scholarship of Imperial College London.
The work of AAT   is supported by the STFC grant ST/T000791/1.

\appendix 

\section{Regularized  Green's   functions  and  coincident limits \la{apG}} 

\subsection{Scalar  in $\AdS_{d+1}$ \la{apG0}}

For   a scalar field    in Euclidean $\AdS_{d+1}$   space  subject to Dirichlet boundary conditions 
 one has (see, e.g.,  \ci{DHoker:2002nbb}))
% with coordinates $\w^{\mu}$  with the action %=(z, w^{i})$ we have
\ba
&\qquad \qquad  S = \frac{1}{2}\int d^{d+1}\w\, \sqrt{g}\, (\nabla_{\mu}\phi\nabla^{\mu}\phi+m^{2}\phi^{2}), \qquad \qquad 
  \\
&\la{a1} \langle \phi(\w)\, \phi(\w') \rangle = G_\phi(\w,\w') \  , \qquad \qquad  (- \nabla^2 + m^2) G_\phi(\w, \w') = \delta(\w,\w') \  , \\
% , \qquad \Delta_{\pm} = \frac{d}{2}\pm\sqrt{\frac{d^{2}}{4}+m^{2}},
\la{a2}
&G_\phi(\w, \w') = \frac{\Gamma(\Delta)}{2^{\Delta}\pi^{d/2}(2\Delta-d)\Gamma(\Delta-\frac{d}{2})}\frac{1}{(u+2)^{\Delta}}
{}_{2}F_{1}\Big(\Delta, \Delta-\frac{d}{2}+\frac{1}{2}, 2\Delta-d+1, \frac{2}{u+2}\Big), \\
&\Delta  = \frac{d}{2}+\sqrt{\frac{d^{2}}{4}+m^{2}}, \qquad\ m^2= \Delta (\Delta-d) \ , \qquad   u = \frac{  (z-z')^2+ (w^v-w'^v)^2}{2z z'} \ . \la{a4}
\ea
Here  $u$ is the chordal  distance  which in \rf{a4} is  specified to the Poincare  coordinates, 
$ds^2_{\AdS_{d+1}} = g_{\m\n} (\w) d\w^\m d\w^\n= { 1\ov z^2} ( dz^2 + dw^v dw^v)$. 

To evaluate $G_\phi(\w, \w') $ in  the coincident point limit $u\to 0$ we use that 
\be
\la{4.5}
_{2}F_{1}(a,b,c; 1) = \frac{\Gamma(c)\Gamma(c-b-a)}{\Gamma(c-b)\Gamma(c-a)}, \qquad \Re(c-a-b)>0, \ \ c\neq 0, -1, -2, \dots,
\ee
and analytically continue in $d$ to extend this relation 
 beyond the above conditions on $a,b,c$ parameters.
We then find that 
\be
\la{a3}
 G_\phi(\w,\w) \equiv \gs_\phi = \frac{1}{(4\pi)^{(d+1)/2}}\frac{\Gamma(\frac{1}{2}-\frac{d}{2}) \, \Gamma(\Delta)}{\Gamma(1-d+\Delta)} \ , \qquad \ \ \ 
\del_\m  G_\phi(\w,\w')  \Big|_{\w=\w'} =0 \ . 
\ee
Considering the second derivative of $G_\phi$  we  conclude that   in dimensional regularization 
\ba
& \partial_{\mu}\partial'_{\nu}G_\phi(\w,\w') \Big|_{\w=\w'}  \equiv  g_{\m\n} \rhg_\phi  \ , \ \ \ \ \ \ \ \ 
\rhg_\phi =
\frac{1}{2(4\pi)^{(d+1)/2}}\frac{\Gamma(-\tfrac{1}{2}-\tfrac{d}{2})\, \Gamma(\Delta+1)}{\Gamma(-d+\Delta)} \ , \la{a5} \\
&
\la{a6} \rhg_\phi =  -\frac{\Delta(\Delta-d)}{d+1}\,  \gs_\phi  = - {m^2 \ov d+1}   \gs_\phi  \ , \ \ \ \ \ \ \ \ \ 
\rh_\phi \equiv   \rhg_\phi   +   {m^2 \ov d+1}   \gs_\phi=0 \ .
%(\gs_\phi)\indices{_{\mu}^{\mu}} = -\Delta(\Delta-d)\, \gs_\phi.
\ea
We assume that the mass  $m$ is fixed, i.e.    does not depend on $d$.
In the case of the  scalar $x^i$ in \rf{36}  with $m^2=3$ in the $p=2$ case  (corresponding to $\Delta=3$  for $d=p$) one finds 
setting $d=2-2\eps, \ \eps\to 0 $ 
\be
\gs_{x} = -\frac{1}{2\pi}+\frac{3-4\LDR}{8\pi}\, \eps+\dots, \qquad 
\rhg_x = {1\ov 2\pi}  + \frac{1  +  4\LDR}{8\pi}\, \eps+\dots, \qquad \rh_x=0 \ ,    \qquad  \LDR \equiv  \log(\pi e^{\gamma_{\rm E}}) \ . \la{a8}
\ee
For the massless  fluctuations $y^a$   we get %Fluctuations in $S^{n}$ are always massless, so $\Delta=d=2$.
%The expansion of propagator is 
\be\la{a9}
\gs_{y} = -\frac{1}{4\pi}-\frac{\LDR}{4\pi}\, \eps+\cdots, \qquad \qquad \rh_y= 0 \ . 
\ee

\iffa
\ba
\la{610}
& \langle\theta\, \bar\theta\rangle = -\G_{*}\sss, \qquad \langle \mk D_{\alpha}\theta\, \bar\theta\rangle = -\G_{\alpha}\ssss ,\qquad
 \langle\mk D_{\alpha}\theta\,  \mk D_{\beta}\theta\rangle = \big(\tgt\, \G_{\alpha\beta}+\bgt\, g_{\alpha\beta}\big)\, \G_{*}\, C^{-1} \ , 
\ea
\fi

%%%%%%%%%%%%%%%%%%%%%%%%%

\subsection{Spinor  in $\AdS_{d+1}$ \la{apG1}}

Let us  consider  a  Euclidean Dirac  operator  with a generalized mass term proportional to a matrix $\bar \gamma$ commuting with Dirac matrices\foot{Below   we  assume for definiteness that $m >0$.}   %(now \underline{commuting} with kinetic term)
\be
\la{413}
\slashed{\mk D} = \slashed{\nabla}+m\, \bar\gamma\,, \qquad \qquad \slashed{\nabla} =\g^\alpha\nabla_\alpha 
\,,\qquad\qquad   \bar\gamma^{2}=1, \ \ \ \ \ \ [\gamma^{\a},\bar\gamma]=0\ .
\ee
The  corresponding Green's function is (see, e.g., \ci{Kawano:1999au,Basu:2006ti}; cf. \rf{a2})
\ba
G(\w,\w') &= \frac{1}{2^{m+(d+3)/2}\pi^{d/2}}\frac{\Gamma(m+\frac{d+1}{2})}{\Gamma(m+\frac{1}{2})} %\frac{1}{\sqrt{zz'}}
\frac{1}{(u+2)^{m+(d+1)/2}} %\lp
%\ \ \ \ \ \ \ \times
 \Big[\bar\gamma(\slashed{\w}\gamma_{0}+\gamma_{0}\slashed{\w}')\, {\rm F}_{1}(u)+(\slashed{\w}-\slashed{\w}')\, {\rm F}_{2}(u)\Big], 
\no \\
{\rm F}_{1}(u) &= {}_{2}F_{1}(m+\tfrac{d+1}{2}, m, 2m+1, \tfrac{2}{u+2}), \qquad
{\rm F}_{2}(u) = {}_{2}F_{1}(m+\tfrac{d+1}{2}, m+1, 2m+1, \tfrac{2}{u+2}). \la{414}
\ea
%Notice minus sign in both terms in second line of (\ref{4.14}). This gives
Similar  expression will apply  in the case of   the fermionic variable $\theta$  in the M2  brane action  
where (cf. \rf{37},\rf{610})
\be \la{455}
 \g_\alpha \to \Gamma_\a, \qquad \ \  \ \bar\gamma \to  \Gamma_{*}  \ , \qquad \ \ \ \   m \to m_f= \te {3\ov 2} \ . 
\ee 
In this case we find  %explicitly 
\ba
\la{416}
G_\theta (\w,\w) \equiv   & -\sss\, \Gamma_{*} \ , \ \ \  
\qquad 
\sss = - \frac{1}{2^{d+1}\pi^{(d+1)/2}}\ \frac{\Gamma(\frac{1-d}{2})\Gamma(\frac{1+d}{2}+m_f)}{\Gamma(\frac{1-d}{2}+m_f)}\ , \\
\la{417}
&\nabla_{\alpha}G_\theta(\w, \w')\Big|_{\w=\w'} = \frac{m_f}{d+1} \G_{\alpha} \sss\ .
\ea
Thus  in dimensional regularization  where  $d=2-2\eps$    and $\G_{\alpha} \G^\alpha = d+1$ 
%so that in full d-dimensions, we have Dirac equation
\be\la{444}
\ssss\equiv - (d+1)^{-1}  \big( \slashed{\nabla}+m_f\G_{*}\big)G_\theta(\w,\w')  \Big|_{\w=\w'} =0\  .
\ee
The expansion of $\sss$  for $\eps\to 0$    and  $m=\frac{3}{2}$ is  given by (cf. \rf{a8})
\be\la{447} 
\sss = \frac{1}{2\pi}+\frac{1+2\LDR}{4\pi}\, \eps+\dots\ .
\ee
In dimensional reduction regularization that we used in  section 3 
  the Dirac algebra is  assumed to be done in 3 dimensions  so that  $\G^{\alpha}\G_{\alpha}=3$. 
Defining   $\mk D_{\alpha}$ as in \rf{37} 
\be
\la{4.20}
\mk D_{\alpha} \equiv  \nabla_{\alpha}+\tfrac{1}{2}\G_{*}\G_{\alpha} = \nabla_{\alpha}+\tfrac{1}{3}m_f \G_{*}\G_{\alpha} \ , 
\ee
we then get  (cf. \rf{610},\rf{611})
\ba
\mk D_{\alpha}
G_\theta(\w,\w')\Big|_{\w=\w'} = &-\G_\a \ssss \ , \qquad 
%= \frac{\sss}{d+1}m\G_{\alpha}+\frac{m}{3}\G_{*}\G_{\alpha}(-\sss\,\G_{*}) 
 \qquad 
\slashed{\mk D}G_\theta(\w,\w')\Big|_{\w=\w'} = - 3\, \ssss \ ,  \la{a199}\\
& \qquad \ssss =  \frac{d-2}{3(d+1)}\, m_f\, \sss=\OO(\eps) \ . \la{a19} 
%m\frac{d-2}{d+1}\,\sss.
\ea

\subsection{$\zeta$-function  and heat kernel regularization \la{apG2}}

In the spectral $\zeta$-function regularization   we fix $d=2$, i.e.  consider all  fields  defined in  Euclidean 
 AdS$_3=\mb H^{3}$.
Given 
 a differential operator $\Delta$ with a positive discrete
  spectrum (with no zero modes), the corresponding 
  Green's function and heat kernel are defined as ($\delta(\s,\s')=\frac{1}{\sqrt g}\delta(\s-\s')$)
\be\la{57}
\Delta\, G(\s,\s') = \delta(\s,\s'), \qquad K(t; \s,\s') = \langle \s| e^{-t \Delta}|\s'\rangle,\qquad \langle \s|\s'\rangle = \delta(\s,\s')\ . 
\ee
For  a complete set of eigenfunctions $f_{n}(\s)$ satisfying 
\be\la{58}
\Delta f_{n}(\s) = \l_{n}f_{n}(\s), \qquad \sum_{n}f_{n}(\s)f_{n}^{\dagger}(\s') = 
\delta(\s,\s'), \qquad \int d^3\s\sqrt{g}\ f_{n}^{\dagger}(\s)f_{n}(\s) = \delta_{nm}\ ,
\ee
the corresponding spectral $\z$-function  is  defined as 
\be\la{59}
\zeta(s; \s,\s') = \sum_{n}\frac{f_{n}(\s)f_{n}^{\dagger}(\s')}{\l_{n}^{s}} = \frac{1}{\Gamma(s)}\int_{0}^{\infty}dt\, t^{s-1}\, K(t; \s,\s').
\ee
This gives
\be\la{60}
G(\s,\s') = \zeta(1; \s,\s'), \qquad\qquad  \Delta\, G(\s,\s') =\delta(\s,\s')=  \zeta(0; \s,\s').
\ee
If the spectrum is continuous, i.e.  $\l_n \to \l(\nu)$,  one may define the spectral measure  $\mu(\nu)$  as  
\be\la{61}
\sum_{n}\to \mc N\ \int_{0}^{\infty}d\nu\, \mu(\nu).
\ee
%\paragraph{Scalar}
For a massive scalar  in $\mb H^{3}$   one finds 
% the trace of the heat kernel and the zeta function are \cite{Camporesi:1991nw,Camporesi:1993mz,Camporesi:1994ga}
\ba
 \la{887} 
 &\Delta  = -\nabla^2 + m^2 \ , \qquad \ \  \qquad \mu(\nu) = \nu^{2}, \qquad \l(\nu) = \nu^{2}+1 + m^{2} \ , \\
&K (t; \s,\s ) = \frac{1}{2\pi^{2}}\int_{0}^{\infty}d\nu\, \mu(\nu)\, e^{-t\l(\nu)}= {1\ov (4 \pi t)^{3/2}}   e^{-t (m^2+1) }
\ , \la{62}
\\
&\zeta(s) = % {1\ov \G(s) } \int^\infty_0  dt   \, t^{s-1} \,   K(t; \s, \s) = 
 \frac{1}{2\pi^{2}}\int_{0}^{\infty}d\nu \, \frac{\mu(\nu)}{[\l(\nu)]^{s}} \la{63}
= \frac{1}{8\pi^{3/2}}(1+m^{2})^{\frac{3}{2}-s}\, \frac{\Gamma(s-\frac{3}{2})}{\Gamma(s)}\ , 
\\ % the  value of the  scalar propagator at coincident points is 
\la{68}
&G(\s,\s)  = \zeta(1) = -\frac{1}{4\pi}\sqrt{1+m^{2}}, \qquad \qquad  \delta(\s,\s)= \zeta(0) = 0\ . %v4
\ea
To find the  twice differentiated propagator at coinciding points we note that 
\be\la{701}
\partial_{\mu}\partial'_{\nu}G(\s,\s') = -\nabla_{\mu}\nabla_{\nu}G(\s,\s') = \tfrac{1}{3}g_{\mu\nu}\big[-m^{2}G(\s,\s')+\delta(\s,\s')\big] \ , 
\ee
 so that   using  \rf{68} we get 
%When $\s'\to \s$ we replace $\frac{1}{\sqrt g}\delta(\s,\s)$ by $\zeta_{B}(0,m^{2})$ (which is zero for scalar). Thus,
\be\la{702} 
\partial_{\mu}\partial'_{\nu}G(\s,\s')|_{\s=\s'} = \frac{1}{12\pi}m^{2}\sqrt{1+m^{2}}\, g_{\mu\nu} \ .
\ee
In the case  of the squared spinor operator in \rf{4.20}   defined on $\mb H^{3}$ one finds \cite{Camporesi:1992tm}\foot{Here  with  hermitian 
 $\G_\a$   one has   $\slashed{\mk D}^{\dagger} =- \G^\a \del_\a + ...$.} % {\bf signs ?!}
\ba \la{703} 
& \qquad \qquad \Delta= \slashed{\mk D}\slashed{\mk D}^{\dagger}\ , \qquad 
\qquad \slashed{\mk D} = \slashed{\nabla}+m\, \G_{*} \ , \\
 \la{704} 
&\mu (\nu) = \nu^{2}+\tfrac{1}{4}, \qquad\qquad  \l(\nu) = \nu^{2}+m^{2},\qquad  
K (t; \s,\s ) = {1\ov (4 \pi t)^{3/2}}   e^{-t m^2 } \, (1 + \ha  t  )\ , 
\\ \la{706}
&\zeta(s) = \frac{m^{1-2s}}{16 \pi^{3/2}}(s-\tfrac{3}{2} +2m^{2})\frac{\Gamma(s-\frac{3}{2})}{\Gamma(s)} \ , 
\qquad \zeta(1) =- \frac{4m^{2}-1}{16\pi m}, \qquad\qquad  \zeta(0)=0\ .
\ea
The  Green's function  for  the operator $\slashed{\mk D}$ defined on the fermion $\theta$   is  (cf. \rf{60})
\be\la{709}
\tG(\s,\s') = \slashed{\mk D}^{\dagger }\zeta(1; \s,\s')\,,\quad  {\mk D}_\a \tG(\s,\s') = \tfrac{1}{3}\G_\a \,  \zeta(0; \s,\s') \ , 
 \quad \slashed{\mk D}\tG(\s,\s') = \zeta(0; \s,\s')= \delta_\theta (\s,\s')  \ .
\ee
Since  on a symmetric space the coincident limit is independent of the point on the manifold,
\ba
    &\tG(\sigma, \sigma) =m\, \Gamma_{*}\zeta(1) \ , \qquad
    \mathfrak{D}_{\alpha}\tG\Big|_{\s=\s'} = \tfrac{1}{3}\gamma_{\alpha} \zeta(0)=0 \ , \qquad 
    %&  \Df_{[\alpha}\Df_{\beta]}\tG\Big|_{\s=\s'}  = - \tfrac{1}{3}\gamma_{\alpha\beta}\G_* \zeta(0) =0     \ , 
    \qquad  \mathfrak{D}_{\alpha} \mathfrak{D}'_{\beta}   \tG\Big|_{\s=\s'}=0 \ . \label{B.30} 
%    & \zeta_{F}(0) = 0 \ , \quad \zeta_{F}(1) = \frac{1}{4\pi}\big( \frac{1}{4m} - m\big) \ ,
\ea
% where in \rf{B.30} we  used the integrability property of the Killing spinor derivative. 
%Also,  from \rf{709}   we find $ \mathfrak{D}_{\alpha} \mathfrak{D}'_{\beta}   G\Big|_{\s=\s'}=0$. 
As a result, 
\be\la{707}
\tG(\s,\s) \equiv - \G_* \rG_{\theta} \ ,  \qquad \ \ \     \rG_{\theta} = - m\,  \zeta(1)=   \frac{4m^{2}-1}{16\pi }\ , \qquad 
\rG_{\theta}\Big|_{m={3\ov 2} } = {1\ov 2 \pi} \ . 
\ee
Let us note  that if instead  of the $\zeta$-function regularization one   uses  the  heat kernel  (or ``proper-time'') cutoff
 so that 
 \be \la{93}
G(\s,\s')=\int_{\epsilon}^{\infty}dt \, K(t; \s,\s') \ , \qquad \delta (\s,\s) = K(\epsilon; \s,\s') \ , \qquad \ \ \  \epsilon \equiv \Lambda^{-2} \to 0 \ , \ee
then  in the above bosonic and fermionic cases we find, respectively,  
\begin{align}
 &\delta(\s,\s) = \frac{1}{(4\pi)^{3/2}} \La^3 -\frac{m^2+1}{(4\pi)^{3/2}} \Lambda +\OO(\La^{-1})
%\frac{(m^2+1)^2\epsilon^{1/2}}{16\pi^{3/2}}-\frac{(m^2+1)^3\epsilon^{3/2}}{48\pi^{3/2}} +\mathcal{O}\big(\epsilon^{5/2}\big) \ , \\
    \ , \qquad 
    G(\sigma, \sigma) = \frac{1}{4\pi^{3/2}}\La -\frac{\sqrt{m^2+1}}{4\pi}+\OO(\La^{-1}) \ ,\la{109} \\
    &\delta_\theta(\s,\s)  = \frac{1}{8 \pi ^{3/2} }\La^3 +\frac{1-2 m^2}{16 \pi ^{3/2}}\La +\OO(\La^{-1}) \ , \qquad 
   \rG_\theta (\s,\s)  =  %m\Gamma_{*}\Big[
   - \frac{m}{4 \pi ^{3/2} }\La + \frac{4 m^2-1}{16 \pi  } + \OO(\La^{-1}) %v4
   %\Big]
    \ . \no
    %-\frac{\left(1-2 m^2\right) \epsilon^{1/2}}{8 \pi ^{3/2}}-\frac{m^2 \left(m^2-1\right) \epsilon^{3/2}}{24 \pi ^{3/2}}+\mathcal{O}\left(\epsilon^{5/2}\right)\Big)
\end{align}
The resulting expectation  value of the quartic Lagrangian   given by the sum of  \rf{77},\rf{613},\rf{617}   then contains power 
divergences
\begin{align}\la{a40}
    \langle {L}_{4}\rangle
    = \frac{3}{32 \pi ^3 }\La^6
    - \frac{27}{32 \pi ^3 }\La^4
    + \frac{7}{8 \pi ^{5/2}}\La^3
    - \frac{7}{128 \pi ^3  }\La^2
    - \frac{13}{8 \pi ^{5/2} }\La
    + \frac{3241}{1536 \pi ^3}  + \OO(\La^{-1}) \ . 
\end{align}
These  may be cancelled  by a contribution of an ultralocal measure  in the M2 brane path integral  that may be contributing  in the heat kernel
regularization as here the ``$\delta(0)$'' terms  are non-vanishing.

\section{M2 brane action in \adsss         \la{apM}}

Here  we shall review the  structure of the BST action \cite{Bergshoeff:1987cm,Bergshoeff:1987qx}  
in \adsss   \ci{deWit:1998yu,Claus:1998fh}  using its  supercoset construction. 

\subsection{Supercoset relations \la{apM1}} %definitions}

We shall assume  the  Euclidean signature, i.e.  AdS$_7= \hyp^7$.
This maximally supersymmetric 11d background has  the superisometry algebra
$\mathfrak{osp}(7,1\,|\,4)$  with  the  even part  $  \mathfrak{so}(7,1)\oplus \mathfrak{usp}(4)$ % $so(7,1) \oplus so(5)$
  and the 
odd  part represented by  32  supercharges transforming in the bi-spinor  representation  of the bosonic  groups.
The corresponding commutation relations may be written as\foot{We  use the notation 
$[\cdot\ , \cdot]$ for the graded commutator 
$
    [U, V] = UV - (-1)^{|U| |V|}VU , \ \  U,V \in \mathfrak{\mathfrak{g}}  , 
$
satisfying the graded Jacobi identity:
$
    [U, [V, W]] = [[U, V], W] +(-1)^{|U| |V|}[V, [U, W]]  
$. 
%In the 11d spinor notation, the structure constants of the  superalgebra can be specified as\footnote{
%The relative coefficients in $[Q, \bar{Q}]$ follow from the graded Jacobi identity.
 The commutator $[Q, \bar{Q}]$ can be also written in the 11d supergravity notation: \
$
    [Q, \bar{Q}] = -2\Gamma^{A}P_{A} +\frac{1}{144}\big{(}\Gamma^{ABCDEF}{\cal F}_{ABCD}+24\Gamma_{AB}{\cal F}^{ABEF} \big{)}M_{EF} ,
$
where  ${\cal F}_{4} = 6\mathrm{r}^{4}\text{vol}_{S^{4}}$. %SKV3
We assume that the radius of AdS$_7$ is 1  and the radius of $S^4$ is $\rr=\ha$ (cf. \rf{452}).
While we formally keep track of the dependence on $\rr$,    the \adsss  background is a 11d supergravity solution and thus the 
M2 brane action is  consistent ($\kappa$-symmetric) only for $\rr=\ha$.}
\begin{align}
     &[M_{rs}, M_{tu}] = \delta_{st}M_{ru} + ... %-\delta_{bd}M_{ac}+\delta_{ad}M_{bc}-\delta_{ac}M_{bd}  
     , \qquad 
    [P_{r}, M_{st}] = \delta_{rs}P_{t}-\delta_{rt}P_{s} , \qquad [P_{r}, P_{s}] = M_{rs}   ,  \no \\
    &\te [M_{ab}, M_{cd}] = \delta_{bc}M_{ad} + ... %.-\delta_{bd}M_{ac}+\delta_{ad}M_{bc}-\delta_{ac}M_{bd}
      , \qquad [P_{a}, M_{bc}] = \delta_{ab}P_{c}-\delta_{ac}P_{b}  , \qquad [P_{a}, P_{b}] = -\frac{1}{\mathrm{r}^2}M_{ab}  ,\no \\
    &\te [P_{r}, Q] = -\frac{1}{2}\Gamma_{r}\Gst Q \ , \quad [P_{a}, Q] = -\frac{1}{2\mathrm{r}}\Gamma_{a}\Gst Q \ , \quad 
    [P_{r}, \bar{Q}] = \frac{1}{2}\bar{Q}\Gst\Gamma_{r} \ , \quad [P_{a}, \bar{Q}] =\frac{1}{2\mathrm{r}}\bar{Q}\Gst\Gamma_{a} \ ,  \nonumber \\
    &\te [M_{AB}, Q] = -\frac{1}{2}\Gamma_{AB}Q \ , \quad [M_{AB}, \bar{Q}] = \frac{1}{2}\bar{Q}\Gamma_{AB}  \ , \qquad \bar{Q} = Q^{T} C \ ,\nonumber \\
    &[Q, \bar{Q}] = -2 \Gamma^{A}P_{A} + \Gst\big{(}\Gamma^{rs}M_{rs}-\tfrac{1}{\mathrm{r}}\Gamma^{ab}M_{ab}\big{)} \label{A12} \  . %v4
\end{align}
where $A,B=0, ..., 10$  and $r,s,...=0, ..., 6$; \  $a,b, ...=7, ..., 10$. 
% $\hyp^7$ radius $L_{\hyp^{7}} = 1$ and  a factor $ \mathrm{r} = \frac{1}{2}$ was introduced to account for the $S^{4}$ radius. We define:
Let us define the supercoset element   %($g_{0}(X)$ is the bosonic part) 
\begin{align}\la{b1}
    g(X, \theta) = g_{0}(X)\exp(\Q) \ , \qquad \ \ \ \Q = \bar{\theta}Q = \bar{Q}\theta\ , \qquad  \ \ \ \   X^A= (X^r,X^a) \ . 
\end{align}
%%%%%%%%%%%%%%%%%%%%%%%%%%%
%To construct the supercoset we use the exponential parametrisation for the fermions:
%and define 
The left-invariant Maurer-Cartan one-form    can be decomposed as:
\begin{align}
   \te  g^{-1}dg = L^{A}P_{A}+\frac{1}{2}L^{AB}M_{AB}+\bar{Q}L \ , \qquad L^{A} = (L^{r}, L^{a}) \ , \qquad L^{AB} = (L^{rs}, L^{ab}) \ ,\la{B3}
\end{align}
where $(L^{A}, L)$ satisfy the structure equations:
\begin{align}
    &\te dL^{A} + L^{AB}\wedge L^{B} =  \bar{L} \wedge \Gamma^{A} L \ ,\la{A16}  \\ &\te
     dL +\frac{1}{4}L^{AB} \wedge \Gamma_{AB}L + \frac{1}{2}L^{r}\wedge\Gamma_{*}\Gamma_{r}L+\frac{1}{2\mathrm{r}}L^{a}\wedge\Gamma_{*}\Gamma_{a}L = 0 \ \label{A17} . %v4
\end{align}
Using \rf{b1}  the  1-form in \rf{B3}   may be represented as 
%To determine the Maurer-Cartan form explicitly, we use the differential of the exponential map:
\begin{align}
\te    g^{-1}dg = e^{-\ad\Q}(g_{0}^{-1}dg_{0}) + \frac{1- e^{-\ad \Q}}{\ad \Q}\, d\Q % \nonumber \\
    =g_{0}^{-1}dg_{0} -(\ad \mathrm{Q}/2)\frac{\sinh^2 (\ad \frac{ \mathrm{Q}}{2})}{({\ad} \frac{\mathrm{Q} }{2})^2}D\mathrm{Q} + \frac{\sinh (\ad \mathrm{Q})}{\ad \mathrm{Q}}D\mathrm{Q} \ ,\la{aa16}
\end{align}
where $\ad(\Q) = [\Q, \cdot]$ and $D$ denotes the Killing spinor derivative defined  as\foot{Note that 
 the Killing spinor  on $\hyp^7\times S^4$ can be obtained using the adjoint action (see also  \cite{Alonso-Alberca:2002wsh}):
$
    \Q_{\kil} %=\bar{\epsilon}Q= \ad_{g_{0}^{-1}}(\Q) 
    = \bar{\epsilon}(g_{0}^{-1}Qg_{0})  .
$}
\begin{align}
    D\Q = d\Q +[g_{0}^{-1}dg_{0}, \Q] \ .
\end{align}
The bosonic 1-form expanded in the generators may be written as  %Let us  also introduce $(E^{A}, \Omega^{AB}$)  appearing in  the bosonic part of the one-form % Maurer-Cartan form:
\begin{align}
    g_{0}^{-1}dg_{0} = E^{A}P_{A} + \tfrac{1}{2}\Omega^{AB}M_{AB} \ , \qquad   E^A=(E^r, E^a), \ \ \  \Omega^{AB}= (\Omega^{rs},\Omega^{ab})\ . 
\end{align}
Explicitly,  we find %in exponential fermionic coordinates for the superalgebra \eqref{A12} the Killing spinor derivative is:
\begin{align}
    D\Q = \bar{Q}(D\theta) \ ,\ \ \  \qquad D\theta = d\theta +\tfrac{1}{4}\Omega^{AB}\Gamma_{AB}\theta +\tfrac{1}{2}\Gamma_{*}\big{(}E^{r}\Gamma_{r}+\tfrac{1}{\mathrm{r}}E^{a}\Gamma_{a}\big{)}\theta \ . %v4
\end{align}
Let us  also define  the  fermion bilinear  $\mathcal{M}^2$ as 
\ba
&\ad^2\Q(\tilde{\Q}) = [\bar{\theta}Q, [\bar{\theta}Q, \bar{Q}\tilde{\theta}]] = \bar{Q}\mathcal{M}^2\tilde{\theta} \ , \qquad \qquad 
\Q =\bar \theta {Q}, \ \ \  \ \ \ \tilde{\Q} = \bar{Q}\tilde \theta\ , \\ 
    &\mathcal{M}^2 = \Gamma_{*}\big{(}\Gamma_{r}\theta\, \bar{\theta}\Gamma^{r}+\tfrac{1}{\mathrm{r}}\Gamma_{a}\theta\,\bar{\theta}\Gamma^{a} \big{)}-\tfrac{1}{2}\big{(}\Gamma_{rs}\theta\, \bar{\theta}\Gamma^{rs}-\tfrac{1}{\mathrm{r}}\Gamma_{ab}\theta\, \bar{\theta}\Gamma^{ab} \big{)}\Gamma_{*} \ . \la{ab16} 
\ea
Then from \rf{aa16} we get 
\begin{align}
    g^{-1}dg = g_{0}^{-1}dg_{0} &+4 \bar{\theta}\Gamma^{A}\big{(}\frac{\sinh^2 \frac{\mathcal{M}}{2}}{\mathcal{M}^2}D\theta \big{)}P_{A}+\bar{Q} \big{(}\frac{\sinh\mathcal{M}}{\mathcal{M}}D\theta \big{)} \nonumber \\ 
    &-2\bar{\theta}\Gamma_{*}\Gamma^{rs}\big{(}\frac{\sinh^2 \frac{\mathcal{M}}{2}}{\mathcal{M}^2}D\theta \big{)}M_{rs}+\frac{2}{\mathrm{r}}\bar{\theta}\Gamma_{*}\Gamma^{ab}\big{(}\frac{\sinh^2 \frac{\mathcal{M}}{2}}{\mathcal{M}^2}D\theta \big{)}M_{ab} \ .\la{b12}
\end{align}
Let  us  split the $\hyp^7$  indices as $r=(\hat \a, i)$ where $\hat \a$  is the AdS$_3$ tangent-space  index and $i$ labels the  4 transverse directions. 
In the  static gauge  we  set %where $X^{\hat a} =\s^^{\hat a}$
(using parametrization that   makes  the  transverse $ SO(4)\times SO(4)$ symmetry manifest)
\begin{align} \label{A.24}
    X = (\sigma^{\alpha}, x^{i}, y^{a}) \ , \qquad 
     g_{0}(X) = g_{\hyp^3}(\sigma)\exp\Bigl(\frac{\mathrm{arctanh} \ \mx/2}{\mx/2}x^{i}P_{i} \Bigr) \exp \Bigl(\frac{\arctan \my/(2\mathrm{r})}{\my/(2\mathrm{r})}y^{a}P_{a} \Bigr) \ , 
\end{align}
where $\mx = \sqrt{x^{i}x_{i}},\ \my  = \sqrt{y^{a}y_{a}}$ and $g_{\hyp^3}(\sigma)$ denotes any parametrisation of a totally geodesic $\hyp^3\subset \hyp^7$. The $\hyp^7\times S^4$ metric is determined by the bosonic  1-form  $E^{A}$
 and has the same form as in  \cite{Drukker:2020swu} (cf. \rf{3399},\rf{357}) 
 %as in  of the Maurer-Cartan form, similar to the conventions used in  \cite{Drukker:2020swu}:
\begin{align} \label{A24}
    &g^{-1}_{\hyp^3}dg_{\hyp^3} = e^{\halpha}P_{\halpha}+\omega^{\halpha\hbeta}M_{\halpha\hbeta} \ , \quad E^{\halpha} = \frac{(1+\tfrac{1}{4}x^2)}{(1-\tfrac{1}{4}x^2)}e^{\halpha} \ , \quad E^{i} = \frac{dx^{i}}{1-\tfrac{1}{4}x^2} \ , \quad E^{a} = \frac{dy^{a}}{1+\tfrac{1}{4\mathrm{r}^2}y^2}  \ , \nonumber  \\
    &%ds^2_{11} = ds^2_{\hyp^7} +ds^2_{S^{4}} \ , \quad
     ds^2_{\hyp^7} = \frac{(1+\tfrac{1}{4}x^2)^2}{(1-\tfrac{1}{4}x^2)^2}ds^2_{\hyp^3} + \frac{dx^{i}dx_{i}}{(1-\tfrac{1}{4}x^2)^2} \ , \qquad ds^2_{S^{4}} = \frac{dy^{a}dy_{a}}{(1+\tfrac{1}{4\mathrm{r}^2}y^2)^2} \ .
\end{align}
Here $e^{\halpha}= e^\halpha_\alpha d\s^\alpha$  and $\omega^{\halpha\hbeta}$   correspond to the  3-bein and 
 spin connection of $\hyp^3$. % ($\halpha,\hbeta$  are  tangent space indices). 
  For example, in the upper half-plane  parametrization of  $\hyp^3$  ($\sigma^\a=(\s^0,\s^1,\s^2)= (z, w^v)$)
 \begin{align}
    &ds^2_{\hyp^3} = \frac{1}{z^2}\big{(}dz^2 + dw^v dw^v \big{)} \ , % \quad \sigma^{\halpha} = (\sigma, \  \vec\tau) \ , \nonumber \\
    \qquad
    e^{\halpha} = \frac{1}{z }d\sigma^\a  \ , \qquad \omega^{\halpha\hbeta} = \frac{1}{z} \big{(} \delta^{\alpha}_{0}d\sigma^{\beta} - \delta^{\beta}_{0}d\sigma^{\alpha} \big{)} \ .
\end{align}
The spin connection $\Omega^{AB}$ is explicitly given by $ \Omega^{\halpha \hbeta} = \omega^{\halpha \hbeta}$ and 
\begin{align}\la{bb16}
   \Omega^{\halpha i} = - \Omega^{i\halpha}= \frac{e^{\halpha}x^{i}}{1-\tfrac{1}{4}x^2} \ , \qquad \Omega^{ij} = % -\frac{1}{2}(x^{i}E^{j}-x^{j   }E^{i}) = 
    -\tfrac{1}{2}\frac{x^{i}dx^{j}-x^{j}dx^{i}}{1-\tfrac{1}{4}x^2} \ , \qquad % \nonumber \\
    \Omega^{ab} %= \frac{1}{2\mathrm{r}^2}(y^{a}E^{b}-y^{b   }E^{a}) 
    = \tfrac{1}{2\mathrm{r}^2}\frac{y^{a}dy^{b}-y^{b}dy^{a}}{1+\tfrac{1}{4\mathrm{r}^2}y^2} \ .
\end{align}
%To compute the correlators, we use upper half-plane coordinates on $\hyp^3$:

\subsection {Expansion  of the M2 brane  action  \la{apM2}  }

The BST  action for the supermembrane in $\hyp^7\times S^{4}$ can be written as \cite{deWit:1998yu} (cf. \rf{31},\rf{3200})
\ba
   & S = \TT_2 \big( \int d^3 \sigma \, \sqrt { h} \ +\ i \int_{M_{4}}\mathcal{H}_{4}\big)  \ ,
   \qquad \ \ \   h_{\alpha\beta} = L^{A}_{\alpha}L^{A}_{\beta} \ , \la{bb17}\\
   &\mathcal{H}_{4} = \tfrac{1}{8\mathrm{r}}\epsilon_{abcd}L^{a}\wedge L^{b}\wedge L^{c} \wedge L^{d} -\tfrac{1}{2}\bar{L}\wedge \Gamma_{AB}L \wedge L^{A}\wedge L^{B} ,
\ea
where $L^{A}_{\alpha} = L^{A}_M \del_\a X^M $ and $\mathcal{H}_{4}$ is a closed 4-form  which is  integrated over a 4d manifold that  has the \adst  world volume as its boundary.\foot{Its closure  follows from \eqref{A16},\eqref{A17} for $\mathrm{r}=\frac{1}{2}$ combined with the 11d Fierz identity
(here $p,q,...$ are spinor indices):

$
    \bar{L}\wedge \Gamma_{AB}L\wedge \bar{L}\wedge\Gamma^{A}L= 0  \leftrightarrow (\Gamma^{A})_{(pq}
    (\Gamma_{AB})_{uv)} = 0 .$}
    The action   is invariant   under the 
the $\kappa$-symmetry % is a local fermionic symmetry of the GS action. From the supercoset perspective, 
which   can be represented as a right action on the supercoset  element
\begin{align}
    g(X, \theta) = g_{0}(X) \exp(\Q) \ \mapsto\ g_{0}(X)\exp(\Q) \exp(\delta \bar{\theta}Q) \ \label{A29} . 
\end{align}
The corresponding vector field $\Xi$ generating a diffeomorphism on the coset satisfies\footnote{A vector field $\Xi$ defines a contraction of the same parity as $\Xi$, i.e., for a one-form $\alpha$: $\iota_{\Xi}\alpha = \langle \Xi, \alpha \rangle,$ and $
   \iota_{\Xi}(\alpha\wedge\beta) = \iota_{\Xi}\alpha \wedge\beta +(-1)^{|\Xi|\cdot|\alpha|}\alpha \wedge \iota_{\Xi}\beta  .
$
%Such definition, in particular, implies the Cartan formula for the Lie derivative $\mathcal{L}=d \iota_{\xi}+\iota_{\xi}d $.
}
\begin{align}
    \iota_{\Xi}(g^{-1}dg) = g^{-1}\delta g = \delta_\kappa\bar{\theta}\, Q \ , \qquad %\nonumber \\
    \iota_{\Xi}\bar{L} = \delta_\kappa\bar{\theta} \ , \qquad \iota_{\Xi}L^{A} = 0 \ , \qquad \iota_{\Xi}L^{AB}  = 0 \ .
\end{align}
Explicitly, the  transformation under  the $\kappa$-symmetry   may be represented as  %For the chosen conventions, it then follows that the GS action is invariant under the local $\kappa$-symmetry of the form:
\begin{align}
   &\qquad \qquad \delta_{\kappa}\theta = \mathbf{P}\kappa  \ , \qquad \ \ \  \mathbf{P}=\tfrac{1}{2}\big{(}\id+\mathbf{\Gamma} \big{)} \ , \\
   & \mathbf{\Gamma} \equiv   \frac{i}{3!\sqrt{h}}\epsilon^{\alpha\beta\gamma}L^{A}_{\alpha}L^{B}_{\beta}L^{C}_{\gamma}\Gamma_{ABC} \ , \qquad \mathbf{\Gamma}^2 =\id \ ,\qquad  \quad h^{\alpha\beta}L^{A}_{\alpha}\Gamma_{A}\mathbf{\Gamma} = \frac{i}{2!\sqrt{h}}\epsilon^{\alpha\beta\gamma}L^{A}_{\beta}L^{B}_{\gamma}\Gamma_{AB}  \ .
\end{align}
To expand the  action  in powers  of $\theta$ it is 
 convenient to  do a rescaling %rescale the fermions with parameter $s$:
\begin{align}\la{b23}
   \theta \rightarrow s \theta \ , \qquad L_s\equiv L(s \theta)  \ , \qquad  L^{A}_{s}\equiv  L^A(s\theta) 
    \ .
\end{align}
It then follows, by considering the  special case of \eqref{A29} with $\delta\theta = \delta s\,  \theta$ and using that $L_{s=0} = 0$, that the WZ term in \rf{bb17} can be written as: 
\begin{align}\la{bb24} %SKV3
    & \int_{M_{4}} \mathcal{H} = {6}\rr^{4} \int_{M_{4}} \mathrm{vol}_{S^4}
     - \int d^3\sigma \, \epsilon^{\alpha\beta\gamma}\int_{0}^{1}ds \, \bar{\theta}\Gamma_{AB}L_{\alpha \, s}\  L^{A}_{\beta \, s}L^{B}_{\gamma \, s} \ .
\end{align}
%where $ \mathrm{vol}_{S^4}$ is the volume form on the unit $S^4$.
%%%%%%%%%%%%%%%%%%%%%%%%%%%%%%
Imposing the $\kappa$-symmetry gauge as in \rf{214}, i.e. 
\begin{align} \label{b24}
    \Pc \theta = 0 \ , \qquad \qquad \Pc = \tfrac{1}{2}\big{(}\id + \Gamma\big{)} \ , \qquad \ \ \Gamma = i \G_{012} \ , 
\end{align}
the expansion of the   basic 1-forms in \rf{b23}    in terms of $\theta$  takes the form
\begin{align}
    L^{A}_{s} = E^{A}+s^2\bar{\theta}\Gamma^{A}D\theta+\tfrac{1}{12}s^4 \bar{\theta}\Gamma^{A}\mathcal{M}^2D\theta+\mathcal{O}(s^6) \ , \qquad \ \ \ L_{s} = s D\theta+\tfrac{1}{6}s^3\mathcal{M}^2D\theta+\mathcal{O}(s^5) \ , 
\end{align}
where   $\mathcal{M}^2$ was defined in \rf{ab16}. As a result,  the induced metric and the integrand in the WZ term in \rf{bb24} take the form 
%The expansion of the induced metric and the 3-form in the WZ term can then be written as:
\begin{align}
    &h_{\a\b}=L^{A}_{\alpha}L^{A}_{\beta} \te = (E^{A}_{\alpha}+\bar{\theta}\Gamma^{A}D_{\alpha}\theta+\frac{1}{12}\bar{\theta}\mathcal{M}^2D_{\alpha}\theta+\dots)(E^{A}_{\beta}+\bar{\theta}\Gamma^{A}D_{\beta}\theta+\frac{1}{12}\bar{\theta}\mathcal{M}^2D_{\beta}\theta+\dots) \ , \\
    &\bar{\theta}\Gamma_{AB}L_{s}\wedge L^{A}_{s}\wedge L^{B}_{s}  = 
     \bar{\theta}\Gamma_{AB}(sD\theta+\tfrac{1}{6}s^3\mathcal{M}^2D\theta) \wedge(E^{A}+s^2\bar{\theta}\Gamma^{A}D\theta) \wedge (E^{B}+s^2\bar{\theta}\Gamma^{B}D\theta) + ... \no 
\end{align}
It is convenient to decompose the operators acting on spinors as % into the parts commuting and anticommuting with $\Gamma$:
\begin{align}
    &D =D^{+} + D^{-} \ , \quad \Gamma D^{+}= D^{+}\Gamma \ , \quad \Gamma D^{-} = -D^{-}\Gamma \ ,  \\
    &\te D^{+} = d+\frac{1}{4}\Omega^{\halpha\hbeta}\Gamma_{\halpha\hbeta}+\frac{1}{2}E^{\halpha}\Gamma_{*}\Gamma_{\halpha}+\frac{1}{4}\Omega^{ij}\Gamma_{ij}+\frac{1}{4}\Omega^{ab}\Gamma_{ab} \ , \no  \\ 
   &\te  D^{-} = \frac{1}{2}\Omega^{\halpha i }\Gamma_{\halpha}\Gamma_{i}+\frac{1}{2}E^{i}\Gamma_{*}\Gamma_{i}+\frac{1}{2\mathrm{r}}E^{a}\Gamma_{*}\Gamma_{a}\no  \ ,\\
    &\mathcal{M}^2 = \mathcal{M}^2_{+} + \mathcal{M}^2_{-} \ , \quad \Gamma\mathcal{M}^2_{+} = \mathcal{M}^2_{+} \Gamma \ , \quad \Gamma\mathcal{M}^2_{-} = -\mathcal{M}^2_{-} \Gamma \ , \la{b20} \\
    &\te \mathcal{M}^2_{+} = \Gamma_{*}\Gamma_{\halpha}\theta\bar{\theta}\Gamma^{\halpha}-\frac{1}{2}\big{(} \Gamma_{\halpha\hbeta}\theta\bar{\theta}\Gamma^{\halpha\hbeta}+\Gamma_{ij}\theta \bar{\theta}\Gamma^{ij} - \frac{1}{\mathrm{r}}\Gamma_{ab}\theta \bar{\theta}\Gamma^{ab}\big{)}\Gamma_{*} \ ,\no  \\
    &\te\mathcal{M}^2_{-} = \Gamma_{*}\big{(}\Gamma_{i}\theta \bar{\theta}\Gamma^{i}+\frac{1}{\mathrm{r}}\Gamma_{a}\theta\bar{\theta}\Gamma^{a} \big{)} -\Gamma_{\halpha}\Gamma_{i}\theta\bar{\theta}\Gamma^{\halpha}\Gamma^{i}\Gamma_{*} \ .\no
\end{align}
Since $\overline{\Gamma\theta} = \overline{\theta}\Gamma$, in the above  $\kappa$-symmetry gauge \rf{b24} 
 one  needs to 
 consider only the terms which involve an even number of operators reversing the $\Gamma$-chirality. Using that $\Gamma\Gamma_{\halpha} =\Gamma_{\halpha}\Gamma$ and $\Gamma\Gamma_{i} = -\Gamma_{i}\Gamma$,  we find %the supercoframe decomposes as:
\begin{align}
    &\te L^{\halpha}_{s} = E^{\halpha} + s^{2}\bar{\theta}\Gamma^{\halpha}D^{+}\theta + \frac{s^4}{12}\bar{\theta}\Gamma^{\halpha}\mathcal{M}^2_{+}D^{+}\theta+\frac{s^4}{12}\bar{\theta}\Gamma^{\halpha}\mathcal{M}^2_{-}D^{-}\theta + \mathcal{O}(s^6) \ , \\
    &\te L^{i, a}_{s} = E^{i, a} +s^2\bar{\theta}\Gamma^{i, a}D^{-}\theta+ \frac{s^4}{12}\bar{\theta}\Gamma^{i, a}\mathcal{M}^2_{+}D^{-}\theta+\frac{s^4}{12}\bar{\theta}\Gamma^{i, a}\mathcal{M}^2_{-}D^{+}\theta + \mathcal{O}(s^6) \ , \\
    &\te L_{s} = L_{s}^{+}+L_{s}^{-} \ , \quad L_{s}^{\pm} = s D^{\pm}\theta+\frac{s^3}{6}\mathcal{M}^2_{+}D^{\pm}\theta + \frac{s^3}{6}\mathcal{M}^2_{-}D^{\mp}\theta+\mathcal{O}(s^5) \ .
\end{align}
The spinor covariant derivative can be further expanded in the  bosonic fluctuations as
\begin{align}
    &\te D^{+} = \Df +\frac{1}{4}x^2 e^{\halpha}\Gamma_{*}\Gamma_{\halpha}-\frac{1}{4}x^{i}dx^{j}\Gamma_{ij}+\frac{1}{4\mathrm{r}^2}y^{a}dy^{b}\Gamma_{ab}+\mathcal{O}(X^4) \ , \qquad \ \ \   \Df = \nabla + \tfrac{1}{2}e^{\halpha}\Gamma_{*}\Gamma_{\halpha}  \ , \\
    &\te D^{-} = \frac{1}{2}e^{\halpha}x^{i}\Gamma_{\halpha}\Gamma_{i}+\frac{1}{2}dx^{i}\Gamma_{*}\Gamma_{i}+\frac{1}{2\mathrm{r}}dy^{a}\Gamma_{*}\Gamma_{a}+\mathcal{O}(X^3) \ , \la{b34}
\end{align}
where $\Df$ is the pull-back of the Killing spinor derivative to the  $\hyp^3$ surface (cf. \rf{37}).

As a  result, using the definitions in \rf{41},\rf{42} we find the following expansion of the M2  brane  Lagrangian to quartic order in the fluctuation fields
(cf. \rf{334},\rf{43}--\rf{45})
\ba
  L &=L_{2}  + {\TT}_2 ^{-1} \big(  L_{\rm 4b}  + L_{\rm 2b, 2f}  + L_{\rm 4f}\big) + ...  \ , \la{473}\qquad \qquad 
    L_{2} = \tfrac{1}{2}g^{\alpha\beta}(h_{\alpha\beta}^{(2b)}+h_{\alpha\beta}^{(2\theta)}),
\\
 L_{\rm 4b} =&\tfrac{1}{4}x^2g^{\alpha\beta}h^{(2x)}_{\alpha\beta} -\tfrac{1}{4\mathrm{r}^2}y^2g^{\alpha\beta}h_{\alpha\beta}^{(2y)}-\tfrac{1}{4}g^{\alpha\delta}g^{\beta\gamma}h^{(2b)}_{\alpha\beta}h^{(2b)}_{\gamma\delta}+\tfrac{1}{8}(g^{\alpha\beta}h_{\alpha\beta}^{(2b)})^2 \nonumber \\
  & \quad \quad \quad +\tfrac{i}{8\mathrm{r}\sqrt{g}}\epsilon^{\alpha\beta\gamma}\epsilon_{ abcd}y^{ a}\partial_{\alpha}y^{ b}\partial_{\beta}y^{ c}\partial_{\gamma}y^{ d} \ , \label{2024} \\
 L_{\rm 2b, 2f} =&\te  \frac{1}{8}g^{\alpha\beta}\big(h^{(2x)}_{\alpha\beta}- 2 h_{\alpha\beta}^{(2y)}\big)\bar\theta\G_{*}\theta
+\frac{3}{8}{x}^{2}g^{\alpha\beta}h_{\alpha\beta}^{(2\theta)}
+\frac{1}{8}g^{\alpha\beta}g^{\gamma\delta}h^{(2b)}_{\alpha\beta}\, h_{\gamma\delta}^{(2\theta)}
-\frac{1}{4}h^{(2b)\alpha\beta}h_{\alpha\beta}^{(2\theta)}\no  \\
&\te  -\frac{1}{2}x^{i}\partial_{\beta}x^{j}\,\bar{\theta}\gamma^{\beta}\Gamma_{ij}\theta+\frac{1}{2\mathrm{r}^2}y^{ a}\partial_{\beta}y^{ b}\, \bar{\theta}\gamma^{\beta}\Gamma_{ a b}\theta+\frac{1}{2} \bar{\theta}\gamma^{\alpha}(\partial_{\alpha}x^{i}\Gamma_{i}+\partial_{\alpha}y^{ a}\Gamma_{ a})x^{j}\,\Gamma_{j}\theta \nonumber  \\
    &\te +\frac{1+\mathrm{r}}{4\mathrm{r}}\partial_{\alpha}x^{i}\partial^{\alpha}y^{ a}\, \bar{\theta}\Gamma_{*}\Gamma_{i}\Gamma_{ a}\theta-\frac{i}{2\sqrt{g}}\epsilon^{\alpha\beta\gamma}\bar{\theta}\gamma_{\beta}(\partial_{\gamma}x^{i}\Gamma_{i}+\partial_{\gamma}y^{ a}\Gamma_{ a})(\partial_{\alpha}x^{j}\Gamma_{j}-\frac{1}{\mathrm{r}}\partial_{\alpha}y^{ a}\Gamma_{ a})\Gamma_{*}\theta
    \nonumber \\
    &\te  -\frac{i}{2\sqrt{g}}\epsilon^{\alpha\beta\gamma}\bar{\theta}(\Gamma_{ij}\partial_{\beta}x^{i}\partial_{\gamma}x^{j}+2\Gamma_{i}\Gamma_{ a}\partial_{\beta}x^{i}\partial_{\gamma}y^{ a}+\Gamma_{ a b}\partial_{\beta}y^{ a}\partial_{\gamma}y^{ b} )\Df_{\alpha}\theta \ , \label{2025} \\
 L_{\rm 4f} =& \te \frac{1}{96}g^{\alpha\beta}\bar\theta\G_{\alpha}\mc M_{+}^{2}\mk D_{\beta}\theta-\frac{1}{16}g^{\alpha\delta}g^{\beta\gamma}h^{(2\theta)}_{\alpha\beta}h^{(2\theta)}_{\gamma\delta}
+\frac{1}{16}(g^{\alpha\beta}h_{\alpha\beta}^{(2\theta)})^{2} \ .  \la{435}
\ea

%%%%%%%%%%%%%%%%%%%%%%%%%%%%%%%%%%%%%%%%
\section{ Fermionic correlators        \la{apF}}

Here we  provide some details of the computation of the 
quadratic and quartic fermionic correlators used in \rf{610}--\rf{617}. 

%\subsection {\la{apF}  } 

We need to evaluate the expectation value of 2-fermion operators of the form 
$
\langle\bar\theta W \theta\rangle, \ \ \langle\bar\theta  W \nabla_{\alpha}\theta\rangle,
$
where $W$ is a string of gamma matrices containing  $\Gamma_{\alpha}$ and also  $\G_i,\G_a$  matrices with transverse indices. 
One requires   $(C W)^{T} = -CW$  for these   combinations  to be non-zero. 
As a result, we may 
apply  the  antisymmetric projection 
\be \la{A.7} 
CW \to \tfrac{1}{2}(CW-(CW)^{T})\ , \qquad \ \ \  W\to \tfrac{1}{2}(W+C^{-1}W^{T}C) \ . \ee
The basic   definitions are (here $p,q$ are spinor indices  and  $\bar\theta = \theta^{T}C$) 
\ba\la{c1}
\langle \theta_{p}(\s) \, \bar\theta^{q}(\s') \rangle =(G_{\theta})_{p}^{\  q}(\s,\s'),\qquad 
 \langle\theta_{p}\theta_{q}\rangle = (G_\theta C^{-1})_{pq} \ , \quad 
\langle\bar\theta W \theta\rangle = -\tr[G_\theta W]\ ,
 % \qquad C^{T}=-C, \qquad \G_{A} = -C^{-1}\G_{A}^{T}C,
%and in order for it to be non zero we require 
\ea
where also $(G_\theta C^{-1})^{T} = -G_\theta  C^{-1}.
$
For  correlators where   the fermionic   field is differentiated we have 
\be
\langle\bar\theta W \del  \theta\rangle =  -\tr[\del G_{\theta}\,  W]\ , \qquad 
\langle \del \theta_{p}\theta_{q}\rangle = (\del G_{\theta} C^{-1})_{pq}.
\ee
We   may consider also  $\langle(\nabla_{\alpha}\theta(\w))_{p}\, \theta_{q}(\w')\rangle 
 = (\nabla_{\alpha}G_\theta(\s,\s')\,  C^{-1})_{pq}$  that leads to \rf{417}. In particular  (cf. \rf{a19}) 
 \ba
\la{B.1}
\langle\bar\theta\slashed{\mk D}\theta\rangle = 3N_{\theta}\, \ssss \ , \qquad \qquad 
\langle\bar\theta\G_*\theta\rangle = N_{\theta}\, \sss \ , 
\ea
where the factor of  $N_{\theta}=16$ comes  from the trace over  the spinor indices ($\theta$ assumed to be subject to the $\kappa$-symmetry gauge  in \rf{b24}). 

For a correlator with  2 derivatives    at  the coincident points   we get (assuming dimensional reduction regularization)
\ba
\la{A.18}
\langle(\nabla_{\alpha}\theta)_{p}\ (\nabla_{\beta}\theta)_{q}\rangle \equiv  (S_{\alpha\beta}C^{-1})_{pq} \ , \qquad 
S_{\alpha\beta} = \tfrac{1}{4}\big[\G_{\alpha\beta}-\tfrac{d(d+1)-4m^{2}}{d+1}g_{\alpha\beta}\big]\, \G_{*}\ . 
\ea
Similarly  (cf. \rf{610})
\ba
&\langle(\mk D_{\alpha}\theta)_{p}(\mk D_{\beta}\theta)_{q}\rangle \equiv (\wh S_{\alpha\beta}C^{-1})_{pq},\qquad \te 
\wh S_{\alpha\beta} = \frac{1}{4}\,\big[ \big(1+\frac{4}{9}\frac{d-5}{d+1}m^{2}\big)\, \G_{\alpha\beta}
-\big(d-\frac{4}{9}\frac{d+4}{d+1}m^{2}\big)g_{\alpha\beta} \big]\G_{*}\ , 
\\ &\qquad \qquad  \wh S_{\alpha\beta} \Big|_{m_f={3\ov 2}}=\te  \frac{d-2}{4(d+1)}\, \big[2\, \G_{\alpha\beta}-(d+2)\,  g_{\alpha\beta} \big]\G_{*}\, \sss \ . \la{c7}
\ea
This   correlator  vanishes in $d=2$  in agreement  with \rf{611}.

For  correlators of  4 fermions  we  have\foot{Here  $W$   and $\wt W$  are two different  combination of $\G$ matrices.} 
\ba
\la{A.13}
& \langle \bar\theta W\theta\ \bar\theta\wt W \theta\rangle =
\tr[W G_{\theta}  ]\tr[\wt W G_{\theta} ]-2\tr[W G_{\theta} \wt W G_{\theta} ] \ , \\
& \langle \bar\theta W \del \theta \ \bar\theta\wt W\theta\rangle = \tr[W \del G_{\theta} ]\tr[\wt W G_{\theta} ]-2\tr[W \del G_{\theta}  \wt W G_{\theta} ] \ , 
\ea
with similar more involved expressions   when  the two  fermions are differentiated.

Below we  will consider the  fermionic correlators at coincident points  assuming  generic regularization. 
Let us start   with 
\ba\la{c10}
\mc Q^{(4)}_{1}\equiv  \langle  \bar\theta\G^{\beta}\mk D_{\alpha}\theta\ \bar\theta\G^{\alpha}\G_{\beta}\G_{*}\theta\rangle
\ea
and first apply the projection  in (\ref{A.7})   using 
$
C^{-1}(\G_{\alpha}\G_{\beta}\G_{*})^{T}C =  \G_{\beta}\G_{\alpha}\G_{*} $, so  that 
$
\G_{\alpha}\G_{\beta}\G_{*}\to \frac{1}{2}(\G_{\alpha}\G_{\beta}\G_{*}+\G_{\beta}\G_{\alpha}\G_{*}) = g_{\alpha\beta}\G_{*}.
$
Then (using the definitions in \rf{610},\rf{416},\rf{417},\rf{a19}) 
\ba
\mc Q^{(4)}_{1} &= \langle  \bar\theta\G^{\beta}\mk D_{\alpha}\theta\ \bar\theta\G^{\alpha}\G_{\beta}\G_{*}\theta\rangle 
= \langle  \bar\theta\slashed{\mk D}\theta\ \bar\theta\G_{*}\theta\rangle \lp
= \tr_{_P}[-3\ssss]\tr_{_P}[\G_{*}(-\G_{*}\sss)]-2\tr_{_P}[-3\ssss\G_{*}(-\G_{*}\sss)] \big]
= 3(N_{\theta}^{2}-2N_{\theta})\,\ssss\,\sss , \la{c11}
\ea
where $\tr_{_P}$   stands for the spinor trace under the  $\kappa$-symmetry projection  in \rf{214},\rf{b24}  so that $\tr_{_P} \id =N_\theta=16$.\foot{It is useful to recall  some standard  Dirac matrix relations:  $\{\G_{\a}, \G_{\b}\} = 2g_{\a\b}$,  $g_{\a}^{\a}=d+1$  so that 

$\qquad 
\G^{\a}\G_{\b}\G_{\a} = -(d-1)\G_{\b}, \ \ 
\G^{\a}\G_{\b\gamma }\G_{\a} = (d-3) \G_{\b\gamma }, \ \ 
\G^{\a}\G_{\b\gamma \delta }\G_{\a} = -(d-5) \G_{\b\gamma\delta}, 
$

$\qquad
\tr[\G^{\a\b}\G_{\a\b}] = -d(d+1)\, \tr\, \id\, , \ \ \ 
\tr[\G^{\a\b\gamma}\G_{\a\b\gamma}] = -d(d^{2}-1)\, \tr\, \id\,  . $
}
 In dimensional reduction regularization with $\ssss$  given by \rf{611}, \rf{a19}   we get 
\ba\te
\mc Q^{(4)}_{1} = \frac{3}{2} \frac{d-2}{d+1}(N_{\theta}^{2}-2N_{\theta})\, {\rm G}_\theta^{2}\ .
\ea
Similarly, we find  (using definitions in \rf{610})
\ba\la{c13}
\mc Q^{(4)}_{2} \equiv \langle  \bar\theta\G^{\alpha}\mk D^{\beta}\theta\ \bar\theta\G_{\beta}{\mk D}_{\alpha}\theta\rangle = 
3\,\ssss^{2}\, N_{\theta}^{2}- (9\ssss^{2}-6{\tgt}\sss+3{\bgt}\sss)\, N_{\theta}.
\ea
In the dimensional reduction regularization  this gives
\ba\la{c14}
\mc Q^{(4)}_{2} =\te \frac{3}{4}\frac{d-2}{(d+1)^{2}} \big[(d-2)N_{\theta}^{2}+(12+4d+d^{2})N_{\theta} \big]\, {\rm G}_\theta^2.
\ea
The  results for other   relevant  quartic correlators  found in an analogous way are summarized   below 
\be
\la{B.25}
\def\arraystretch{1.3}
\begin{array}[t]{lll}
%\toprule
% & \langle\bar\theta\slashed{\mk D}\theta\rangle & 3N_{\theta}\,\ssss  \\
% & \langle\bar\theta\G_{*}\theta\rangle & N_{\theta}\, \sss  \\
%\midrule
\mc Q^{(4)}_{1}= & \langle  \bar\theta\G^{\beta}\mk D_{\alpha}\theta\ \bar\theta\G^{\alpha}\G_{\beta}\G_{*}\theta\rangle
& =3(N_{\theta}^{2}-2N_{\theta})\,\ssss\,\sss \\
%%%%%%%%%%%
\mc Q^{(4)}_{2}= & \langle  \bar\theta\G^{\alpha}\mk D^{\beta}\theta\ \bar\theta\G_{\beta}{\mk D}_{\alpha}\theta\rangle
&=3\ssss^{2}\, N_{\theta}^{2}-(9\ssss^{2}-6{\tgt}\sss+3{\bgt}\sss)\, N_{\theta} \\
%%%%%%%%%%%%
\mc Q^{(4)}_{3}= &\langle  \bar\theta\slashed{\mk D}\theta\ \bar\theta\slashed{\mk D}\theta\rangle 
&=9\ssss^{2}\, N_{\theta}^{2}-(9\ssss^{2}+6{\tgt}\sss+3{\bgt}\sss)\, N_{\theta}\\
%%%%%%%%%%%%%
\mc Q^{(4)}_{4}= & \langle\bar\theta\G^{\beta\l}\G_{*}\mk D^{\alpha}\theta\ \bar\theta\G_{\alpha}\G_{\beta\l}\theta\rangle
& =12\,\ssss\sss N_{\theta} \\
%%%%%%%%%%%%%%%%
\mc Q^{(4)}_{5}= & \langle \bar\theta \G^{ij}\G_{*}\mk D_{\alpha}\theta\, \bar\theta\G^{\alpha}\G_{ij}\theta\rangle
& =72\, \ssss\sss N_{\theta}\\
%%%%%%%%%%%%%%%%
\mc Q^{(4)}_{6}= & \langle \bar\theta \G^{ab}\G_{*}\mk D_{\alpha}\theta\, \bar\theta\G^{\alpha}\G_{ab}\theta\rangle
& =72\, \ssss\sss N_{\theta}
\end{array}
\ee
In the dimensional reduction regularization that gives 
\be
\la{B.26}
\def\arraystretch{1.3}
\begin{array}[t]{lll}
%\toprule
%\mc Q^{(2)}_{1} & \langle\bar\theta\G_{*}\theta\rangle & N_{\theta}\, \sss  \\
%\mc Q^{(2)}_{2} & \langle\bar\theta\slashed{\nabla}\theta\rangle &  -\frac{3m}{d+1}\, N_{\theta}\, \sss \\
%\midrule
\mc Q^{(4)}_{1}= & \langle  \bar\theta\G^{\beta}\mk D_{\alpha}\theta\ \bar\theta\G^{\alpha}\G_{\beta}\G_{*}\theta\rangle
& =\frac{3}{2}\frac{d-2}{d+1}(N_{\theta}^{2}-2N_{\theta})\, {\rm G}_\theta^2 \\
%%%%%%%%%%%
\mc Q^{(4)}_{2}= & \langle  \bar\theta\G^{\alpha}\mk D^{\beta}\theta\ \bar\theta\G_{\beta}{\mk D}_{\alpha}\theta\rangle
&=\frac{3}{4}\frac{d-2}{(d+1)^{2}} [(d-2)N_{\theta}^{2}+(12+4d+d^{2})N_{\theta}]\, {\rm G}_\theta^2 \\
%%%%%%%%%%%%
\mc Q^{(4)}_{3}= &\langle  \bar\theta\slashed{\mk D}\theta\ \bar\theta\slashed{\mk D}\theta\rangle 
&=\frac{3}{4}\frac{(d-2)^{2}}{(d+1)^{2}} [3N_{\theta}^{2}+(d-2)N_{\theta}]\, {\rm G}_\theta^2\\
%%%%%%%%%%%%%
\mc Q^{(4)}_{4}= & \langle\bar\theta\G^{\beta\l}\G_{*}\mk D^{\alpha}\theta\ \bar\theta\G_{\alpha}\G_{\beta\l}\theta\rangle
& =6\,\frac{d-2}{d+1}N_{\theta}\, {\rm G}_\theta^2 \\
%%%%%%%%%%%%%%%%
\mc Q^{(4)}_{5}= & \langle \bar\theta \G^{ij}\G_{*}\mk D_{\alpha}\theta\, \bar\theta\G^{\alpha}\G_{ij}\theta\rangle
& =36\, \frac{d-2}{d+1}N_{\theta}\, {\rm G}_\theta^2\\
%%%%%%%%%%%%%%%%
\mc Q^{(4)}_{6}= & \langle \bar\theta \G^{ab}\G_{*}\mk D_{\alpha}\theta\, \bar\theta\G^{\alpha}\G_{ab}\theta\rangle
& =36\, \frac{d-2}{d+1}N_{\theta}\, {\rm G}_\theta^2
\end{array}
\ee

\newpage

\small 
\bibliography{BT-Biblio}
\bibliographystyle{JHEP-v2.9}
\end{document}